\documentclass[prd,%
preprint,%
tightenlines,superscriptaddress,showpacs,%
nofootinbib,eqsecnum,amsfonts,amsmath,amssymb]{revtex4}
\usepackage{bm}
\usepackage{bm,graphics,graphicx,epsfig,color}
\usepackage{subfigure}
\usepackage{amsfonts}
\usepackage{amsopn}
\usepackage{amssymb}
\usepackage{pxfonts}
\usepackage{rotating}
\usepackage{setspace}
\usepackage{amssymb}
\usepackage{lscape}
\usepackage{bm}
\usepackage[varg]{txfonts}

\newcommand{\bs}{\begin{subequations}}
\newcommand{\es}{\end{subequations}}
\newcommand{\beq}{\begin{equation}} \newcommand{\eeq}{\end{equation}}
\newcommand{\bea}{\begin{eqnarray}} \newcommand{\eea}{\end{eqnarray}}
\newcommand{\ba}{\begin{array}} \newcommand{\ea}{\end{array}}
\newcommand{\beqn}{\begin{eqnarray*}}
\newcommand{\ii}{\mathrm{i}}
\newcommand{\eeqn}{\end{eqnarray*}}

     \def\ii{\mathrm{i}}
  
\allowdisplaybreaks
\begin{document}

\title{Tail effects in the third post-Newtonian gravitational wave energy flux \\of
compact binaries in quasi-elliptical orbits}

\author{K G Arun}
\email{arun@iap.fr} \affiliation{Raman Research Institute, Bangalore
560 080, India} 
\affiliation{${\mathcal{G}}{\mathbb{R}}
\varepsilon{\mathbb{C}}{\mathcal{O}}$, Institut d'Astrophysique de Paris
--- C.N.R.S., 98$^{\text{bis}}$ boulevard Arago, 75014 Paris, France}
\affiliation{LAL, Universit{\'e} Paris-Sud, IN2P3/CNRS, Orsay, France}
\author{Luc Blanchet} \email{blanchet@iap.fr}
\affiliation{${\mathcal{G}}{\mathbb{R}}
\varepsilon{\mathbb{C}}{\mathcal{O}}$, Institut d'Astrophysique de Paris
--- C.N.R.S., 98$^{\text{bis}}$ boulevard Arago, 75014 Paris, France}
\author{Bala R Iyer} \email{bri@rri.res.in} \affiliation{Raman Research
Institute, Bangalore 560 080, India} \author{Moh'd S S
Qusailah}\email{mssq@rri.res.in}
 \affiliation{Raman Research Institute, Bangalore
560 080, India}
\affiliation{University of Sana, Yemen}
\begin{abstract}
The far-zone flux of energy contains hereditary (tail) contributions
that depend on the entire past history of the source. Using the
multipolar post-Minkowskian wave generation formalism, we propose and
implement a semi-analytical method in the frequency domain
 to compute these contributions from
the inspiral phase of a binary system of compact objects moving in
quasi-elliptical orbits up to third post-Newtonian (3PN) order. 
The method explicitly uses
the quasi-Keplerian representation of elliptical orbits at 1PN order
and exploits the doubly periodic nature
of the motion to average
the 3PN fluxes over the binary's orbit. Together with the
instantaneous (non-tail) contributions evaluated in a companion paper,
it provides crucial inputs for the construction of ready-to-use
templates for compact binaries moving on quasi-elliptic orbits, an
interesting class of sources for the ground based gravitational
wave detectors such as LIGO and Virgo as well as space
based detectors like LISA.
\end{abstract}
\date{\today} \pacs{04.30.-w, 04.25.-g}

\maketitle

\section{Introduction}\label{secI}
The gravitational-wave (GW) energy flux from a system of two point
masses in elliptic motion in the leading quadrupolar approximation
(Newtonian order) was first obtained by Peters \&
Mathews~\cite{PM63,Pe64}. Using the first post-Newtonian (1PN) order
 quasi-Keplerian representation
of the binary's orbit~\cite{DD85}, Blanchet \& Sch\"afer~\cite{BS89}
computed the 1PN corrections to the above result (confirming earlier
work by Wagoner \& Will~\cite{WagW76}).\footnote{As usual the $n$PN
order refers to the post-Newtonian terms of order $(v/c)^{2n}$ where
$v$ denotes the typical binary's orbital velocity and $c$ is the speed
of light.} Using the generalized quasi-Keplerian representation of the
2PN motion~\cite{DS88,SW93,Wex95}, Gopakumar and Iyer~\cite{GI97}
extended these results to 2PN order and computed the `secular'
evolution of orbital elements under 2PN gravitational
radiation-reaction (4.5PN terms in the equations of motion).  These
constitute one of the basic inputs for gravitational wave phasing of
binaries in quasi-eccentric orbits in the adiabatic approximation. All
these works above relate to the \textit{instantaneous} terms in the
phasing of gravitational waves.

The multipole moments describing GWs emitted by an isolated system do
not evolve independently. They couple to each other and with
themselves, giving rise to non-linear physical effects.  Consequently,
starting at relative 1.5PN order, the above instantaneous terms in the
flux must be supplemented by the contributions arising from these
non-linear multipole interactions. The leading multipole interaction
is between the mass quadrupole moment $I_{ij}$ and the mass monopole
$M$ or  Arnowitt, Deser and Misner (ADM) mass. It is associated with the non-linear effect of tails
at order 1.5PN, and is physically due to the backscatter of linear
waves from the space-time curvature generated by the total mass
$M$. Tails imply a non-locality in time since they are described as
integrals depending on the history of the source from the remote past
to the current retarded time. They are thus appropriately referred to
as \textit{hereditary} contributions by Blanchet \&
Damour~\cite{BD88,BD92} -- terms non-local in time depending on the
dynamics of the system in its entire past~\cite{BD92}. The most
detailed study of tails in this context~\cite{B98tail,B98quad} is
based on the multipolar post-Minkowskian formalism~\cite{BD86,B87}. Up
to 3PN order the hereditary terms comprise the dominant
quadratic-order tails, 
the quadratic-order memory integral~\cite{Chr91,WiW91,Th92,BD92,ABIQ04} 
and  the cubic-order tails. 
The latter \textit{cubic} ``monopole-monopole-quadrupole''
 interaction can be called ``tails of
tails'' of GWs (see~\cite{B98tail,B98quad} for earlier references to
the general topic of tails). In this paper we set up a general
theoretical framework to compute the hereditary contributions for
binaries moving in elliptical orbits and apply it to evaluate {\it
all} the tail contributions contained in the 3PN accurate GW energy
flux.

For the instantaneous terms in the energy flux, explicit closed form
analytical expressions can be given in terms of dynamical variables
related to relative velocity and relative separation. Consequently,
these expressions can be conveniently averaged in the time domain over
an orbit using their quasi-Keplerian representation. For the
hereditary contributions on the other hand one can only write down
formal analytical expressions as integrals over the past.  More
explicit expressions in terms of the dynamical variables \textit{a
priori} require a model of the binary's orbital evolution in the past
to implement the integration. In general, one can show~\cite{BD92}
that the past-influence of tails decreases with some kernel $\propto
1/(t-t')^2$ where $t$ is the current time and $t'$ the integration
time in the past. Thus the ``remote-past'' contribution to the tail
integrals is negligible. More precisely, it was shown~\cite{BS93} that
the contribution due to the past of the tail integral is $\mathcal{O}(\xi_\mathrm{rad}\ln\xi_\mathrm{rad})$ where
$\xi_\mathrm{rad}\equiv\dot{\omega}/\omega^2$ is the adiabatic
parameter associated with the binary's inspiral due to radiation
reaction, which is of order 2.5PN. Consequently, the tail integrals
may be evaluated using standard integrals for a \textit{fixed}
non-decaying circular orbit and errors due to inspiral by gravitation
radiation reaction are at least 4PN order~\cite{BS93}.

In the circular orbit case, with the above simplified model of binary inspiral
one can work directly in the time domain. For instance, the hereditary
terms in the flux were computed up to 3.5PN~\cite{B98tail,B98quad}
while those in the GW polarisations could be obtained up to
2.5PN~\cite{BIWW96,ABIQ04}. In the elliptic orbit case on the other hand the
situation is more involved. Even after using the quasi-Keplerian
parametrization, one cannot perform the integrals in the time domain
(as for the circular orbit case), since the multipole moments have a
more complicated dependence on time and the integrals are not
analytically solvable in simple closed forms. By working in the Fourier
domain Ref.~\cite{BS93} computed the hereditary tail terms at 1.5PN
for elliptical orbits using the lowest order Keplerian representation.

In the present investigation we tackle the terms at orders 2.5PN and
3PN and we need to go beyond the (Newtonian) Keplerian representation
of the orbit to a 1PN quasi-Keplerian representation. Here we
encounter two further complications. Firstly, the 1PN parametrization
of the binary~\cite{DD85} involves three kinds of eccentricities
($e_r$, $e_t$ and $e_\phi$). More seriously, at 1PN order the
periastron precession effect appears in the problem and one has to
contend with two times scales: the orbital time scale and the
periastron precession time scale. These new features are to be
properly accounted for in the calculations to extend the Fourier
method of Ref.~\cite{BS93}. This strategy has been proposed and used
earlier in computing the instantaneous terms in the GW
polarizations from binaries on elliptical
orbits~\cite{D83,GI02,DGI04}. We shall adapt these features here to
treat the more involved hereditary contribution to the total energy
flux.

Following Ref.~\cite{BS93}, we express all the multipole moments
needed for the hereditary computation at Newtonian order as discrete
Fourier series in the mean anomaly of motion $\ell$. However, for the quadrupole moment $I_{ij}$
needed beyond the lowest Newtonian order, the ``doubly periodic''
nature of the motion
needs to be crucially incorporated. The evaluation of the Fourier
coefficients is done numerically based on a series of combinations of
Bessel functions. All tail terms at 2.5PN and 3PN are computed to
provide the ``enhancement factors'' (functions of eccentricity
playing a role similar to the classic Peters \& Mathews~\cite{PM63}
enhancement factor) for binaries in elliptical orbits
at the 2.5PN and 3PN orders. The present work extends results for hereditary
contributions at 1.5PN~\cite{BS93} for elliptical orbits to 2.5PN and
3PN orders. The 3PN hereditary contributions comprise the tail-of-tail
terms and are also extensions of~\cite{B98quad,B98tail} for circular orbits
to the elliptical case.\footnote{Recall that tails are not just
mathematical curiosities in general relativity but facets that should
show up in the GW signals of inspiraling compact binaries and be
decoded by the detectors Virgo/LIGO and
LISA~\cite{BSat94,BSat95,AIQS06a,AIQS06b}.}

Combining the \textit{hereditary} contributions computed in this paper
with the \textit{instantaneous} contributions computed in the
companion paper~\cite{ABIQ07} will yield the complete 3PN energy flux,
generalizing the circular orbit results at 2.5PN~\cite{B96} and
3PN~\cite{BIJ02,BFIJ02,BDEI04} to the elliptical orbit case. The final expressions
represent GWs from a binary evolving adiabatically under gravitational
radiation reaction, including precisely up to 3PN order the effects of
eccentricity and periastron precession during epochs of inspiral when
the orbital parameters are essentially constant over a few orbital
revolutions. It thus represents the first input to go towards the full
\textit{quasi-elliptical} case, namely the evolution of the binary in
an elliptical orbit under gravitational radiation reaction.

Recently, Damour, Gopakumar \& Iyer~\cite{DGI04} proposed an analytic
method based on an improved ``method of variation of constants'' to
construct high accuracy templates for the GW signals from the inspiral
phase of compact binaries moving in quasi-elliptical orbits. The three
time scales, respectively related to orbital motion, orbital
precession and radiation reaction, are handled without the usual
approximation of assuming adiabaticity relative to the radiation
reaction time scale. The explicit results of the above
treatment~\cite{DGI04} relate to ``Newtonian'' radiation reaction
(2.5PN terms in the equations of motion).  It leads to post-adiabatic
(fast) oscillations resulting in amplitude corrections at order 2.5PN
beyond the secular terms.  More recently this work has been
extended~\cite{KG06} to 1PN radiation reaction (3.5PN terms in the
equations of motion).\footnote{For circular orbits, \textit{secular}
evolution of the phase, computed in the adiabatic approximation, is
known up to 3.5PN order~\cite{BFIJ02,BDEI04}.}

The paper is organized as follows: In Section~\ref{secII} we review
the solution of the equations of motion of compact binaries and
discuss its important properties relevant for this present
work. Section ~\ref{secIII} provides the Fourier decomposition of
multipole moments and its use in averaging the energy
flux. Section~\ref{secIV} provides the computations of all the tail
contributions  whose numerical implementation is elaborated in
Section~\ref{secV}. The complete 3PN contributions are exhibited in
Section~\ref{secVI} together with relevant checks. The paper ends with
an Appendix listing the Fourier coefficients of the required Newtonian
moments in terms of the Bessel functions.
\section{Solution of the equations of motion of compact binaries}\label{secII}
\subsection{Doubly-periodic structure of the solution}
\label{secIIA} 
In this work and the next one~\cite{ABIQ07}, we shall often need to
use the explicit solution for the motion of non-spinning compact
binary systems in the post-Newtonian (PN) approximation. We review
here the relevant material we need, which includes the general
``doubly-periodic'' structure of the PN solution, and the
quasi-Keplerian representation of the 1PN binary motion by means of
different types of eccentricities. We closely follow the
works~\cite{D83houches,D83,DD85}.

The equations of motion of a compact binary system up to the 3PN order
admit, when neglecting the radiation reaction term at the 2.5PN order,
ten first integrals of the motion corresponding to the conservation of
energy, angular and linear momenta, and position of the center of
mass~\cite{DJSequiv,ABF01}. When restricted to the frame of the center of mass,
the equations admit four first integrals associated with the energy
$E$ and the angular momentum vector $\mathbf{J}$, given at 3PN order
by Eqs.~(4.8)--(4.9) of Ref.~\cite{BI03CM}.

The motion takes place in the plane orthogonal to
$\mathbf{J}$. Denoting by $r=\vert\mathbf{x}\vert$ the binary's
orbital separation in that plane, and by
$\mathbf{v}=\mathbf{v}_1-\mathbf{v}_2$ the relative velocity, we find
that $E$ and $\mathbf{J}$ are functions of $r$, $\dot{r}^2$, $v^2$ and
$\mathbf{x}\times\mathbf{v}$ (we are employing for definiteness the
harmonic coordinate system of~\cite{BI03CM}\footnote{All calculations
in this paper will be done at the relative 1PN order, and at that
order there is no difference between the harmonic and ADM
coordinates.}), and depend on the total mass $m=m_1+m_2$ and reduced
mass $\mu=m_1m_2/m$. We adopt polar coordinates $r$, $\phi$ in the
orbital plane, and express $E$ and the norm $J=\vert\mathbf{J}\vert$,
thanks to $v^2=\dot{r}^2+r^2\dot{\phi}^2$, as some explicit functions
of $r$, $\dot{r}^2$ and $\dot{\phi}$. The latter functions can be
inverted (by means of straightforward PN iteration) to give
$\dot{r}^2$ and $\dot{\phi}$ in terms of $r$ and the constants of
motion $E$ and $J$. Hence,
\bs
\label{eom}
\bea \dot{r}^2 &=& \mathcal{R}[r;E,J],\\ \dot{\phi} &=&
\mathcal{G}[r;E,J], \eea \es
where the functions $\mathcal{R}$ and $\mathcal{G}$ denote certain
polynomials in $1/r$, the degree of which depends on the PN
approximation in question (it is seventh degree for both $\mathcal{R}$
and $\mathcal{G}$ at 3PN order~\cite{MGS04}). The various coefficients
of the powers of $1/r$ are themselves polynomials in $E$ and $J$, and
also, of course, depend on $m$ and the dimensionless reduced mass ratio
 $\nu\equiv\mu/m$. In
the case of bounded elliptic-like motion, one can prove~\cite{D83}
that the function $\mathcal{R}$ admits two real roots, $r_\mathrm{P}$
and $r_\mathrm{A}$ such that $r_\mathrm{P}<r_\mathrm{A}$, which admit
some non-zero finite Newtonian limits when $c\rightarrow\infty$, and
represent respectively the radii of the orbit's periastron and
apastron. The other roots tend to zero when $c\rightarrow\infty$.

We are considering a given binary's orbital configuration, fully
specified by some given values of the integrals of motion $E$ and $J$.
We no longer indicate the dependence on $E$ and $J$ which is always
implicit in what follows. The binary's orbital period, or time of
return to the periastron, is obtained by integrating the radial motion
as
\beq\label{P}
P = 2 \int_{r_\mathrm{P}}^{r_\mathrm{A}}\frac{d
r}{\sqrt{\mathcal{R}[r]}}.
\eeq
We introduce the fractional angle (\textit{i.e.} the angle divided by
$2\pi$) of the advance of the periastron per orbital revolution,
\beq\label{K}
K = \frac{1}{\pi}\int_{r_\mathrm{P}}^{r_\mathrm{A}}d r
\frac{\mathcal{G}[r]}{\sqrt{\mathcal{R}[r]}},
\eeq
which is such that the precession of the periastron per period is
given by $\Delta\phi=2\pi(K-1)$. As $K$ tends to one in the limit
$c\rightarrow\infty$ (as is easily checked from the Newtonian limit),
it is often convenient to pose $k\equiv K-1$, which will then entirely
describe the \textit{relativistic precession}.

Let us define the mean anomaly $\ell$ and the mean motion $n$ by
\bs\label{elln}\bea \ell &=& n(t-t_\mathrm{P}),\\ n &=&
\frac{2\pi}{P}.  \eea \es
Here $t_\mathrm{P}$ denotes the instant of passage to the periastron.
For a given value of the mean anomaly $\ell$, the orbital separation
$r$ is obtained by inversion of the integral equation
\beq\label{ell}
\ell = n \int_{r_\mathrm{P}}^{r} \frac{d r'}{\sqrt{\mathcal{R}\left[r'\right]}}.
\eeq
This defines the function $r(\ell)$ which is a periodic function in
$\ell$ with period $2\pi$. The orbital phase $\phi$ is then obtained
in terms of the mean anomaly $\ell$ by integrating the angular motion
as
\beq\label{phi}
\phi = \phi_\mathrm{P} + \frac{1}{n} \int_0^{\ell} d
\ell'\,\mathcal{G}\left[r(\ell')\right],
\eeq
where $\phi_\mathrm{P}$ denotes the value of the phase at the instant
$t_\mathrm{P}$. In the particular case of a circular orbit,
$r=\mathrm{const}$, the phase evolves linearly with time,
$\dot{\phi}=\mathcal{G}\left[r\right]=\omega$, where $\omega$ is the
orbital frequency of the circular orbit given by
\beq\label{omega} \omega = K\,n = (1+k)\,n.  \eeq
In the general case of a non-circular orbit it is convenient to keep
the definition of $\omega=K n$ (which will notably be very useful in
the next work~\cite{ABIQ07}) and to explicitly introduce the linearly
growing part of the orbital phase~\eqref{phi} by writing it in the
form
\begin{align}\label{phidecomp}
\phi &= \phi_\mathrm{P} + \omega\,(t-t_\mathrm{P}) +
W(\ell)\nonumber\\ &= \phi_\mathrm{P} + K\,\ell + W(\ell).
\end{align}
Here $W(\ell)$ denotes a certain function which is periodic in $\ell$
(hence, periodic in time with period $P$). According
to~\eqref{phi} this function is given in terms of the mean anomaly
$\ell$ by
\beq\label{W}
W(\ell) = \frac{1}{n} \int_0^{\ell} d
\ell'\,\Bigl[\mathcal{G}\left[r(\ell')\right]-\omega\Bigr].
\eeq
Finally, the decomposition~\eqref{phidecomp} exhibits clearly the
``doubly periodic'' nature of the binary motion, in terms of the mean
anomaly $\ell$ with period $2\pi$, and in terms of the periastron
advance $K\,\ell$ with period $2\pi\,K$.\,\footnote{
Recall, that though standard, the term ``doubly periodic'' may mislead a bit
in that the motion in physical space is not periodic in
 general.
 The radial motion $r(t)$ is periodic with period $P$ while the angular motion
$\phi(t)$ is periodic [modulo $2\pi$] with period $P/k$ where
$k=K-1$. Only when the two period are commensurable, \textit{i.e.}
when $k=1/N$ where $N$ is a natural integer, is the motion periodic
in physical space (with period $N P$).} 
It may be noted that in Refs.~\cite{GI02,DGI04}
the notation $\lambda$ is used; it corresponds to $\lambda=K\,\ell$
and will also occasionally be used here.
\subsection{Quasi-Keplerian representation of the motion 
of compact binaries}\label{secIIB}
In the following we shall also use the explicit solution of the motion
at 1PN order, in the form due to Damour \& Deruelle~\cite{DD85}.  The
solution is given in parametric form in terms of the eccentric anomaly
$u$. Then the radius $r$ and mean anomaly $\ell$ are expressed as
\bs\label{rellkepler}
\begin{align}
r &= a_r\,(1-e_r \cos u),\label{rell}\\ \ell &= u - e_t \sin
u.\label{kepler}
\end{align}
\es
The phase angle $\phi$ is given by (the additive constant
$\phi_\mathrm{P}$ is for convenience set equal to zero)
\beq\label{KV}
\phi = K\,V,\\
\eeq
where the true anomaly $V$ is defined by\footnote{We have denoted the
true anomaly by $V$ rather than by the symbol $v$ of earlier papers to
avoid conflict with the relative speed $v$.}
\beq\label{V} V= 2 \arctan \Biggl[ \Biggl( \frac{ 1 + e_{\phi}}{ 1 -
e_{\phi}} \Biggr)^{1/2} \!\tan \frac{u}{2} \Biggr].  \eeq
In the above, $K$ is the periastron advance given in general terms by
Eq.~\eqref{K}, and $a_r$ is the semi-major axis of the orbit. Note
that there are, in this parametrization at 1PN order, three kinds of
eccentricities $e_r$, $e_t$ and $e_{\phi}$ (labelled after the
coordinates $r$, $t$ and $\phi$). All these eccentricities differ from
one another by 1PN terms, while the advance of the periastron per
orbital revolution appears also starting at the 1PN order. Due to
these features, this representation is referred to as the
``quasi-Keplerian'' (QK) parametrization for the 1PN orbital motion of
the binary. The periodic function $W$ of Eq.~\eqref{W} now reads
\beq\label{WK}
W = K \left( V - \ell\right)\,.\\
\eeq
To close the above solution we need to know the explicit dependence of
the orbital elements in terms of the 1PN conserved energy $E$ and
angular momentum $J$ in the center-of-mass frame (taken as usual per
unit of the reduced mass $\mu$). This is given in
Ref.~\cite{DD85}. Note that the semi-major axis $a_r$ and mean motion
$n$ depend at 1PN order only on the constant of energy through
\bs\label{arn}\bea a_r &=& - \frac{G \,m}{2 E}\left\{ 1+\left(
\frac{7}{2}-\frac{\nu}{2} \right)\frac{E}{c^2}\right\},\\n &=&
\frac{(-2\,E)^{3/2}}{G \,m} \left\{ 1+\left(
\frac{15}{4}-\frac{\nu}{4} \right)\frac{E}{c^2}\right\}.  \eea\es
Posing $h\equiv J/(Gm)$, the 1PN periastron precession simply
reads\footnote{Thus it is sometimes useful to define $k'=k/3$ which
reduces to $1/(c^2 h^2)$ at 1PN order.}
\beq\label{Kexpl} K = 1+\frac{3}{c^2\,h^2}, \eeq
while the three different eccentricities are given by
\bs\label{eretephi}\bea e_r &=& \left\{1 + 2\,E\,h^2\left[ 1+\left(
  -\frac{15}{2}+\frac{5}{2}\nu \right)\frac{E}{c^2}+\frac{-6+\nu}{c^2
    h^2}\right]\right\}^{1/2},\\e_t &=& \left\{1 + 2\,E\,h^2\left[
  1+\left( \frac{17}{2}-\frac{7}{2}\nu
  \right)\frac{E}{c^2}+\frac{2-2\nu}{c^2
    h^2}\right]\right\}^{1/2},\\e_\phi &=& \left\{1 + 2\,E\,h^2\left[
  1+\left( -\frac{15}{2}+\frac{\nu}{2}
  \right)\frac{E}{c^2}-\frac{6}{c^2 h^2}\right]\right\}^{1/2}.\eea \es
Notice the following simple ratios (valid at 1PN order)
\bs\label{ratios}\bea \frac{e_t}{e_r} &=& 1+\left( 8-3\nu\right)
\frac{E}{c^2},\\\frac{e_t}{e_\phi} &=& 1+\left( 8-2\nu\right)
\frac{E}{c^2},\\\frac{e_r}{e_\phi} &=& 1+\nu\frac{E}{c^2}.\eea \es
In the following paper~\cite{ABIQ07} we shall need and use the explicit
solution of the generalized QK binary motion up to 3PN order.
\section{Fourier decomposition of the binary's multipole moments}
\label{secIII} 
\subsection{Peters \& Mathews derivation of the Newtonian energy flux}\label{secIIIA}
The method we shall use in this paper is exemplified by the
computation of the averaged energy flux of compact binaries at
Newtonian order using a Fourier decomposition of the Keplerian
motion~\cite{PM63}. The GW energy flux, say
\beq\label{flux} {\cal{F}} \equiv \left(\frac{d{\cal{E}}}{d
t}\right)^\mathrm{GW} \equiv \left(\int d\Omega\,\frac{d{\cal{E}}}{d
t\,d\Omega}\right)^\mathrm{GW}, \eeq
where $\cal{E}$ is the energy carried in the gravitational waves,
reduces at Newtonian order to the standard Einstein quadrupole
formula\footnote{From now on we set  $c=1$ and $G=1$.}
\beq\label{Fquadorder} {\cal{F}}^{(\mathrm{N})} =
\frac{1}{5}\,\mathop{I_{ij}}^{(3)}{}^{\!\!(\mathrm{N})}(t)
\mathop{I_{ij}}^{(3)}{}^{\!\!(\mathrm{N})}(t), \eeq
where $(\mathrm{N})$ means the Newtonian limit, the superscript $(n)$
refers to differentiation w.r.t time $n$ times, and $I_{ij}^{(\mathrm{N})}$ is the
symmetric-trace-free (STF) quadrupole moment at Newtonian order given
by
\beq\label{IijN} I_{ij}^{(\mathrm{N})} = \mu\,x^{<i}x^{j>}. \eeq
Here $x^i$ is the binary's orbital separation, and the angular
brackets around indices indicate the STF projection:
$x^{<i}x^{j>}\equiv x^{i}x^{j}-\frac{1}{3}\delta^{ij}r^2$. Peters \&
Mathews~\cite{PM63} obtained the expression of the (averaged)
Newtonian flux for compact binaries on eccentric orbits by two
methods. The first method was to take directly the average in time of
Eq.~\eqref{Fquadorder} using the expression~\eqref{IijN} computed for
the Keplerian ellipse; the second method was to decompose the
components of the quadrupole moment into discrete Fourier series using
the known Fourier decomposition of the Keplerian motion (the two
methods, as expected, agreed on the result).

In the second method the quadrupole moment, which is a periodic
function of time at Newtonian order, is thus decomposed into the
Fourier series
\bs\label{IijfourierN}\bea I_{ij}^{(\mathrm{N})}(t) &=&
\sum_{p=-\infty}^{+\infty}\,\mathop{{\cal{I}}}_{(p)}{}_{\!\!ij}^{(\mathrm{N})}\,e^{\ii
\,p\,\ell},\\
\text{with}~~\mathop{{\cal{I}}}_{(p)}{}_{\!\!ij}^{(\mathrm{N})} &=&
\int_0^{2\pi}\frac{d\ell}{2\pi}\,I_{ij}^{(\mathrm{N})}\,e^{-\ii\, p\,\ell},
\eea\es
where $\ell$ is the mean anomaly of the binary motion,
Eq.~\eqref{elln}. Since $I_{ij}^{(\mathrm{N})}$ is real the Fourier
discrete coefficients satisfy
${}_{(p)}{\cal{I}}_{ij}^{(\mathrm{N})}={}_{(-p)}{\cal{I}}_{ij}^{(\mathrm{N})*}$
($*$ denotes the complex conjugate). Inserting
Eqs.~\eqref{IijfourierN} into~\eqref{Fquadorder} we obtain
\beq\label{Fquadfourier} {\cal{F}}^{(\mathrm{N})} =
\frac{1}{5}\sum_{p=-\infty}^{+\infty}\sum_{q=-\infty}^{+\infty} (\ii
p\,n)^3 (\ii q\,n)^3\mathop{{\cal{I}}}_{(p)}{}_{\!\!ij}^{(\mathrm{N})}
\mathop{{\cal{I}}}_{(q)}{}_{\!\!ij}^{(\mathrm{N})}\,e^{{\mathrm i} (p+q)
\ell}.\eeq
Next we perform an average over one period $P$ which means the average
over $\ell=n\,(t-t_\mathrm{P})$ which is easily performed with the
formula
\beq\label{avN}
\langle e^{\ii p \ell}\rangle\equiv \int_0^{2\pi}\frac{d\ell}{2\pi}\,e^{\ii
\,p \,\ell} = \delta_{p,0}.
\eeq
This immediately yields the averaged energy flux in the form of the
Fourier series
\beq\label{avFN} \langle {\cal{F}}^{(\mathrm{N})}\rangle =
\frac{2}{5}\sum_{p=1}^{+\infty}(p\,n)^{6}\,
\vert\!\mathop{{\cal{I}}}_{(p)}{}_{\!\!ij}^{\!\!(\mathrm{N})}\vert^2.
\eeq
Using dimensional analysis (and the known circular orbit limit) this flux is necessarily of the form
\beq
\label{avFNfe} \langle {\cal{F}}^{(\mathrm{N})}\rangle =
\frac{32}{5}\,\nu^2\,\left(\frac{m}{a}\right)^5 f(e), \eeq
where $\nu=\mu/m$ and $a$ is the semi-major axis of the Newtonian
orbit, and the function $f(e)$ is a dimensionless function depending
only on the binary's eccentricity $e$. The coefficient in front
of~\eqref{avFNfe} is chosen in such a way that $f(e)$ reduces to one
for circular orbits, \textit{i.e.} when $e=0$. Thus we have
\bea
\label{PMf} f(e) = \frac{1}{16\,\mu^2\,a^4}\sum_{p=1}^{+\infty} p^6
\,\vert\!\mathop{{\cal{I}}}_{(p)}{}_{\!\!ij}^{\!\!(\mathrm{N})}\vert^2.
\eea
The Fourier coefficients of the quadrupole moment are explicitly
given by Eqs.~\eqref{A3} in the Appendix~\ref{appA} below.
Remarkably this function admits an algebraically closed-form
expression, crucial for the timing of the binary pulsar
PSR~1913+16~\cite{TW82}, and given by
\beq\label{fe}
f(e) = \frac{1+\frac{73}{24}e^2+\frac{37}{96}e^4}{(1-e^2)^{7/2}}.
\eeq
The function $f(e)$ is the Peters \& Mathews~\cite{PM63}
``enhancement'' function, so designated because in the case
of the binary pulsar, which has eccentricity $e=0.617\cdots$, it
enhances the effect of the orbital $\dot{P}$ by a factor $\sim
11.843$. The proof that the series~\eqref{PMf} can be summed up to
yield the closed-form expression~\eqref{fe} is given in the Appendix
of~\cite{PM63}. 
Of course Eq.~\eqref{fe} is in full
agreement with the direct computation of the average performed in the
time domain~\cite{PM63}, \textit{i.e.}
\beq\label{feav} f(e) =
\frac{1}{32\,\mu^2\,a^4\,n^6}\,\langle\mathop{I_{ij}}^{(3)}{}^{\!\!(\mathrm{N})}
\mathop{I_{ij}}^{(3)}{}^{\!\!(\mathrm{N})}\rangle.  \eeq

The method of decomposing the Newtonian moment of compact binaries as
discrete Fourier series was used in Ref.~\cite{BS93} to compute the
tail at the dominant 1.5PN order. To extend this result we need to be
more systematic about the Fourier decomposition of the (not
necessarily Newtonian) source multipole moments.
\subsection{General structure of the Fourier decomposition}\label{secIIIB}
The two sets of source-type multipole moments of the compact binary
system are denoted by $I_L(t)$ and $J_L(t)$ following
Ref.~\cite{B98mult}. Here the multi-index notation means $L\equiv
i_1i_2\cdots i_l$, where $l$ is the number of indices or multipolarity
(which is not to be confused with the mean anomaly $\ell$). In this
Section we investigate the structure of the mass and current moments
$I_L$ and say $J_{L-1}$ (where $L-1\equiv i_1i_2\cdots i_{l-1}$ is
chosen in the current moment for convenience rather than $L$), at any
PN order and for a compact binary system moving on a general
non-circular orbit\footnote{However the intrinsic spins of the compact
objects are neglected, so the motion takes place in a fixed orbital
plane.}.  Their general structure can be written as
\bs\label{ILJLgen1} \bea I_L(t) &=& \sum_{k=0}^{l}
{\cal{F}}_k[r,\dot{r},v^2]\,x^{ <i_1\cdots i_k}v^{i_{k+1}\cdots
i_l>},\label{ILgen1}\\ J_{L-1}(t) &=& \sum_{k=0}^{l-2}
\mathcal{G}_k[r,\dot{r},v^2]\,x^{ <i_1\cdots i_k}v^{i_{k+1}\cdots
i_{l-2}}\varepsilon^{i_{l-1}>ab}x^av^b,\label{JLgen1} \eea \es
where $\varepsilon^{iab}$ is the Levi-Civita symbol (such that
$\varepsilon^{123}=1$), where $x^i=y_1^i-y_2^i$ and $v^i=d
x^i/dt=v_1^i-v_2^i$ denote the relative position and ordinary velocity
of the two bodies (in a harmonic coordinate
system). In~\eqref{ILJLgen1} we pose for instance $x^{i_1\cdots
i_k}\equiv x^{i_1}\cdots x^{i_k}$, and the angular brackets
surrounding indices refer to the usual symmetric-trace-free (STF)
projection with respect to those indices.

Using polar coordinates $r$, $\phi$ in the orbital plane (as in
Sec.~\ref{secIIA}), the above introduced coefficients ${\cal{F}}_k$
and $\mathcal{G}_k$ depend on the masses and on $r$, $\dot{r}$ and
$v^2=\dot{r}^2+r^2\dot{\phi}^2$. For quasi-elliptic motion we can
explicitly factorize out the dependence on the orbital phase $\phi$ by
inserting $x=r\cos\phi$, $y=r\sin\phi$, and
$v_x=\dot{r}\cos\phi-r\,\dot{\phi}\sin\phi$,
$v_y=\dot{r}\sin\phi+r\,\dot{\phi}\cos\phi$. Furthermore, using the
explicit solution of the motion (Sec.~\ref{secIIB}) we can express
$r$, $\dot{r}$ and $v^2$, and hence the ${\cal{F}}_k$'s and
$\mathcal{G}_k$'s, as periodic functions of the mean anomaly
$\ell=n\,(t-t_\mathrm{P})$, where $n=2\pi/P$. We then find that the
above general structure of the multipole moments can be expressed in
terms of the phase angle $\phi$, as the following finite sum over some
``magnetic-type'' index $m$ ranging from $-l$ to $+l$,
\bs\label{ILJLgen2} \bea I_L(t) &=& \sum_{m=-l}^{l}
\,\mathop{\mathcal{A}}_{(m)}{}_{\!\!L}(\ell)\,e^{\ii\,m\,\phi},\\
J_{L-1}(t) &=& \sum_{m=-l}^{l}
\,\mathop{\mathcal{B}}_{(m)}{}_{\!\!L-1}(\ell)\,e^{\ii\,m\,\phi}, \eea
\es
involving some coefficients ${}_{(m)}\mathcal{A}_L$ and
${}_{(m)}\mathcal{B}_{L-1}$ depending on the mean anomaly $\ell$ and
which are complex ($\in\mathbb{C}$). (Some of these coefficients could
be vanishing in particular cases.) The point for our purpose is that
these coefficients are \textit{periodic} functions of $\ell$ with
period $2\pi$. As we can see, the structure of the mass and current
moments $I_L$ and $J_{L-1}$ is basically the same, but their
coefficients ${}_{(m)}\mathcal{A}_L$ and ${}_{(m)}\mathcal{B}_{L-1}$
will have a different parity, because of the Levi-Civita symbol
entering the current moment $J_{L-1}$.

To proceed further, let us exploit the doubly periodic nature of the
dynamics in the two variables $\lambda\equiv K\,\ell$ and $\ell$ (as
reviewed in Sec.~\ref{secIIA}). The phase is given in full generality
by Eq.~\eqref{phidecomp} where we recall that $W(\ell)$ is periodic in
$\ell$. In the following it will be more convenient to single out in
the expression of the phase the purely relativistic precession of the
periastron, namely $\lambda-\ell=k\,\ell$ where $k=K-1$. Inserting the
expression of the phase variable into Eqs.~\eqref{ILJLgen2} yields
many factors which do modify the coefficients of~\eqref{ILJLgen2}, but
in such a way that they remain periodic in $\ell$. Hence we can write
\bs
\label{ILJLgen3}
\bea I_L(t) &=& \sum_{m=-l}^{l}
\,\mathop{{\cal{I}}}_{(m)}{}_{\!\!L}(\ell)\,e^{\ii \,m \,k
\,\ell},\label{ILgen3}\\ J_{L-1}(t) &=& \sum_{m=-l}^{l}
\,\mathop{\mathcal{J}}_{(m)}{}_{\!\!L-1}(\ell)\,e^{\ii \,m \,k
\,\ell},\label{JLgen3} \eea \es
where the coefficients ${}_{(m)}\mathcal{I}_L(\ell)$ and
${}_{(m)}\mathcal{J}_{L-1}(\ell)$ are $2\pi$-periodic. Finally, this
makes it possible to use a discrete Fourier series expansion in the
interval $\ell\in [0,2\pi]$ for each of these coefficients, namely
\bs
\label{decompfourier}
\bea \mathop{{\cal{I}}}_{(m)}{}_{\!\!L}(\ell) &=&
\sum_{p=-\infty}^{+\infty}\,\mathop{{\cal{I}}}_{(p,m)}{}_{\!\!L}\,e^{\ii
\,p\,\ell},\\ \mathop{\mathcal{J}}_{(m)}{}_{\!\!\!L-1}(\ell) &=&
\sum_{p=-\infty}^{+\infty}\,\mathop{\mathcal{J}}_{(p,m)}{}_{\!\!\!L-1}\,e^{\ii
\,p\,\ell}, \eea\es
with inverse relations given by
\bs
\label{invdecompfourier}
\bea \mathop{{\cal{I}}}_{(p,m)}{}_{\!\!L} &=& \int_0^{2\pi}\frac{d
\ell}{2\pi}\,\mathop{{\cal{I}}}_{(m)}{}_{\!\!L}(\ell)\,e^{- \ii
\,p\,\ell},\\ \mathop{{\cal{J}}}_{(p,m)}{}_{\!\!L-1} &=&
\int_0^{2\pi}\frac{d
\ell}{2\pi}\,\mathop{{\cal{J}}}_{(m)}{}_{\!\!L-1}(\ell)\,e^{- \ii
\,p\,\ell}.  \eea\es
This leads then to the following final decompositions of the multipole
moments,
\bs
\label{ILJLfourier}
\bea I_L(t) &=& \sum_{p=-\infty}^{+\infty}\,\sum_{m=-l}^{l}
\,\mathop{{\cal{I}}}_{(p,m)}{}_{\!\!L}\,e^{\ii \,(p+m \,k)
\,\ell},\label{ILfourier}\\ J_{L-1}(t) &=&
\sum_{p=-\infty}^{+\infty}\,\sum_{m=-l}^{l}
\,\mathop{\mathcal{J}}_{(p,m)}{}_{\!\!\!L-1}\,e^{\ii \,(p+m \,k)
\,\ell}.\label{JLfourier} \eea\es
Obviously, since the moments $I_L$ and $J_{L-1}$ are real, their
Fourier coefficients must satisfy
${}_{(p,m)}{\cal{I}}_L={}_{(-p,-m)}{\cal{I}}^*_L$ and
${}_{(p,m)}\mathcal{J}_{L-1}={}_{(-p,-m)}{\cal{J}}^*_{L-1}$. 

The previous decompositions were general, but it is still useful to
introduce a special notation for the particular case of the Newtonian
(N) order, for which the relativistic precession $k$ tends to zero. In
this case we recover the usual periodic Fourier decomposition of the
moments [generalizing Eqs.~\eqref{IijfourierN}], with only one Fourier
summation over the index $p$, so that
\bs
\label{ILJLfourierN}
\bea I_L^{(\mathrm{N})}(t) &=&
\sum_{p=-\infty}^{+\infty}\,\mathop{{\cal{I}}}_{(p)}{}_{\!\!L}^{(\mathrm{N})}\,e^{\ii
\,p \,\ell},\label{ILfourierN}\\ J_{L-1}^{(\mathrm{N})}(t) &=&
\sum_{p=-\infty}^{+\infty}\,\mathop{\mathcal{J}}_{(p)}
{}_{\!\!\!L-1}^{(\mathrm{N})}\,e^{\ii \,p\,\ell}.\label{JLfourierN} \eea
\es
The Newtonian Fourier coefficients are equal to the sums over $m$ of
the doubly-periodic Fourier coefficients in Eqs.~\eqref{ILJLfourier}
when taken in the Newtonian limit, namely
\bs
\label{coeffN}
\bea \mathop{{\cal{I}}}_{(p)}{}_{\!\!L}^{(\mathrm{N})} &=&
\sum_{m=-l}^{l}\,\mathop{{\cal{I}}}_{(p,m)}{}_{\!\!L}^{\!\!(\mathrm{N})},\\
\mathop{\mathcal{J}}_{(p)}{}_{\!\!\!L-1}^{(\mathrm{N})} &=&
\sum_{m=-l}^{l}\,\mathop{\mathcal{J}}_{(p,m)}{}_{\!\!\!L-1}^{\!\!(\mathrm{N})}.
\eea\es
\section{Tail contributions in the flux of compact binaries}
\label{secIV} 
The technique of the previous Section is applied to the computation of
the tail integrals in the energy flux of compact binaries. Although
the computations are effectively done up to the 3PN level, the method
we propose could in principle be implemented at any PN order.
\subsection{Expression of the tail integrals in the 3PN energy flux}\label{secIVA}
As reviewed in the Introduction, the first hereditary term in the
energy flux ${\cal F}$ occurs at the 1.5PN order and is due to GW
tails caused by interaction between the mass quadrupole moment and the
total ADM mass. At the 3PN order, three kinds of hereditary terms
appear: (1) The tails caused by quadratic non-linear interaction
between higher-order multipole moments with the mass; (2) the ``tails
of tails'' due to the cubic non-linear interaction between the tail
itself and the mass; (3) a particular ``tail-squared'' term arising
from self-interaction of the tail\footnote{Recall that the hereditary
character of the non-linear memory
integral~\cite{Chr91,WiW91,Th92,BD92,ABIQ04} is that of a time
anti-derivative in the waveform (\textit{i.e.} the radiative
moments). Thus the non-linear memory becomes instantaneous in the
energy flux, which is made out of time derivatives of the radiative
moments, and will be included into the instantaneous terms computed
in~\cite{ABIQ07}.}.

In the equations to follow, we list the expressions for all these
hereditary tail terms. They are given as non-local integrals over the
source multipole moments of the system $I_{ij}(t)$, $I_{ijk}(t)$,
... and $J_{ij}(t)$, ..., where we use the specific definition of the
PN source moments given in Ref.~\cite{B98mult}. Thus the energy flux
${\cal F}$ defined by Eq.~\eqref{flux} can be split at 3PN order into
\bea
\label{F3PN} {\cal F}^\mathrm{(3PN)} = {\cal{F}}_\mathrm{inst} + 
{\cal{F}}_\mathrm{hered}, 
\eea
where the ``instantaneous'' part, which depends on the source moments
at the same instant (say $t$), reduces at the Newtonian order to the
Einstein quadrupole moment flux ${\cal{F}}^{(\mathrm{N})}$ given by
Eq.~\eqref{Fquadorder}. On the other hand, the ``hereditary'' part
reads
\bea
\label{Fhered} {\cal F}_\mathrm{hered} = {\cal{F}}_\mathrm{tail} + 
{\cal{F}}_\mathrm{tail(tail)} + {\cal{F}}_{\mathrm{(tail)}^2},
\eea
where the quadratic-order tail integrals are explicitly given by (see
Ref.~\cite{BIJ02})\footnote{For convenience we do not indicate the
neglected PN terms, \textit{e.g.} ${\cal{O}}(c^{-n})$. All equations
are valid through the aimed 3PN precision. In the companion
paper~\cite{ABIQ07} we shall restore all powers of $1/c$ (and $G$).}
\begin{align}\label{Ftail}
{\cal{F}}_\mathrm{tail} &=
\frac{4M}{5}\,I_{ij}^{(3)}(t)\,\int_0^{+\infty}d\tau\,I^{(5)}_{
ij}(t-\tau)\biggl[\ln\left(\frac{\tau}{2r_0}\right)
+\frac{11}{12}\biggr]\nonumber\\ &+
\frac{4M}{189}\,I_{ijk}^{(4)}(t)\,\int_0^{+\infty}d\tau\,I^{(6)}_{
ijk}(t-\tau)\biggl[\ln\left(\frac{
\tau}{2r_0}\right)+\frac{97}{60}\biggr]\nonumber\\ &+
\frac{64M}{45}\,J_{ij}^{(3)}(t)\,\int_0^{+\infty}d\tau\,J^{(5)}_{
ij}(t-\tau)\biggl[\ln\left(\frac{
\tau}{2r_0}\right)+\frac{7}{6}\biggr],
\end{align}
while the cubic-order tails (proportional to $M^2$) are
\bs
\label{Ftailtail}\begin{align}
{\cal{F}}_\mathrm{tail(tail)} &=
\frac{4M^2}{5}\,I_{ij}^{(3)}(t)\,\int_0^{+\infty}d\tau\,I^{(6)}_{
ij}(t-\tau)\biggl[\ln^2\left(\frac{
\tau}{2r_0}\right)+\frac{57}{70}\ln\left(\frac{
\tau}{2r_0}\right)+\frac{124627}{44100}\biggr],\\{\cal{F}}_{{\rm
(tail)}^2} &= \frac{4M^2}{5}\left(\int_0^{+\infty}d\tau\,I^{(5)}_{
ij}(t-\tau)\biggl[\ln\left(\frac{
\tau}{2r_0}\right)+\frac{11}{12}\biggr]\right)^2.
\end{align}\es
In these expressions recall that $M$ is the \textit{conserved} mass
monopole or total ADM mass of the source. The first term
in~\eqref{Ftail} is the dominant tail at order 1.5PN while the second
and third represent the sub-dominant tails both appearing at order
2.5PN. The higher-order tails are not given since they are at least at
3.5PN order (see~\cite{B98tail} for their expressions). The two
cubic-order tails given in Eqs.~\eqref{Ftailtail} are both at 3PN
order.

The constant $r_0$ scaling the logarithms in the above tail integrals
has been defined to match with the choice made in the computation of
tails-of-tails in Ref.~\cite{B98tail}. This is the length scale
appearing within the regularization factor $(r/r_0)^B$ used in the
multipolar moment formalism valid for general
sources~\cite{B98mult}. Note that $r_0$ is a freely specifiable
constant entering the relation between the retarded time 
in radiative coordinates 
[used in Eqs.~\eqref{Ftail}-\eqref{Ftailtail}] 
and the corresponding time  in
harmonic coordinates. 
Hence $r_0$ merely relates the origins of time in
the two coordinate systems and is unobservable.

We shall compute all the tail and tail-of-tail
terms~\eqref{Ftail}--\eqref{Ftailtail} [\textit{i.e.}  up to the 3PN
order] averaged over the mean anomaly $\ell$. Together with the
instantaneous terms reported in the next paper~\cite{ABIQ07} we shall
obtain the complete expression of the 3PN energy flux. It is clear
from Eqs.~\eqref{Ftail}--\eqref{Ftailtail} that all the terms
necessitate an evaluation at the relative Newtonian order {\it except}
the mass-type quadrupolar tail term -- first term in~\eqref{Ftail}
-- which must crucially include the 1PN corrections.  We start with
all the terms required at relative Newtonian order and then tackle the
more difficult 1PN quadrupolar tail term.
\subsection{Tails at relative Newtonian order}\label{secIVB}
As a warm up, we consider the mass-type quadrupolar tail term in the
energy flux, the first term in Eq.~\eqref{Ftail}, but given
simply at the relative Newtonian order, namely\footnote{We shall
compute this term at 1PN relative order in Sec.~\ref{secIVD}.}
\beq\label{FquadtailN} \langle{\cal{F}}_{\rm
mass\;quad}^{(\mathrm{N})}\rangle_{\rm tail} = \langle
\frac{4M}{5}\,\mathop{I_{ij}}^{(3)}{}^{\!\!(\mathrm{N})}(t)\,\int_0^{+\infty}d\tau\,
\mathop{I_{ij}}^{(5)}{}^{\!\!(\mathrm{N})}(t-\tau)
\biggl[\ln\left(\frac{\tau}{2r_0}\right) +\frac{11}{12}\biggr]\rangle,
\eeq
where the brackets $\langle\rangle$ refer to the average over the mean
anomaly $\ell$ as defined by Eq.~\eqref{avN}. The
term~\eqref{FquadtailN} was already computed using a Fourier series at
Newtonian order in Ref.~\cite{BS93}; note that the method
of~\cite{BS93} is valid only for periodic motion and thus is
applicable only at the Newtonian level. In this Section we recover the
Newtonian result of~\cite{BS93}.

The Fourier decomposition of the Newtonian quadrupole moment was
already given in general form by Eqs.~\eqref{IijfourierN}. We insert that
decomposition into the flux~\eqref{FquadtailN} and we evaluate the
tail integral by using the fact that if $\ell(t)=n\,(t-t_\mathrm{P})$
corresponds to the current time $t$, then clearly
$\ell(t-\tau)=\ell(t)-n\,\tau$ corresponds to the retarded time
$t-\tau$. Next we perform the average over the current value $\ell(t)$
with the help of the formula~\eqref{avN}. The result is
\beq\label{FtailfourierN} \langle{\cal{F}}_\mathrm{mass\;
quad}^{(\mathrm{N})}\rangle_{\rm tail} =
-\frac{4M}{5}\!\sum_{p=-\infty}^{+\infty}(p\,n)^{8}\,
\vert\!\mathop{{\cal{I}}}_{(p)}{}_{\!\!ij}^{\!\!(\mathrm{N})}\vert^2
\,\int_0^{+\infty}d\tau\,e^{\ii p\,n\,\tau}
\biggl[\ln\left(\frac{\tau}{2r_0}\right)+\frac{11}{12}\biggr].  \eeq
It remains to handle the last factor in~\eqref{FtailfourierN} which is
the tail integral in the Fourier domain, and which is computed using
the closed-form formula
\beq\label{int1}
\int_0^{+\infty}d\tau\,e^{\ii
\,\sigma\,\tau}\ln\left(\frac{\tau}{2r_0}\right) =
-\frac{1}{\sigma}\left[\frac{\pi}{2}\mathrm{sign}(\sigma) +
\ii\Bigl(\ln(2\vert \sigma\vert r_0) + C\Bigr)\right],
\eeq
where $\sigma\equiv p\,n$, $\mathrm{sign}(\sigma)=\pm 1$ and
$C=0.577\cdots$ denotes the Euler constant. Inserting Eq.~\eqref{int1}
into~\eqref{FtailfourierN}, we check that the imaginary parts cancel
out, and the result reduces to
\beq\label{FtailN}
\langle{\cal{F}}_\mathrm{mass\;quad}^{(\mathrm{N})}\rangle_{\rm tail} = \frac{4\pi
M}{5}\sum_{p=1}^{+\infty}(p\,n)^{7}\,
\vert\!\mathop{{\cal{I}}}_{(p)}{}_{\!\!ij}^{\!\!(\mathrm{N})}\vert^2.
\eeq
Observe that the range of $p$'s corresponds to positive frequencies
only. Eq.~\eqref{FtailN} agrees with the result of~\cite{BS93} and can
interestingly be compared with the expression of the Newtonian part of
the averaged flux (quadrupole formula) as given by
Eq.~\eqref{avFN}. Although Eq.~\eqref{FtailN} is expressed in terms of
the relatively simple Fourier series~\eqref{FtailN} [unlike for the
case of the 1PN quadrupole tail in Sec.~\ref{secIVD} which will turn
out to be substantially more intricate], it has to be left in this
form since no analytic closed-form expression can be found for the
infinite sum of these Fourier components~\cite{BS93}. This is in
contrast with the quadrupolar Newtonian flux~\eqref{avFN} which does
admit a closed-form expression [recall Eq.~\eqref{fe}]. In
Sec.~\ref{secV} we shall further proceed following Ref.~\cite{BS93} by
expressing Eq.~\eqref{FtailN} in terms of a new ``enhancement'' factor
depending on the eccentricity and which will be computed numerically.

Let us stress that the result~\eqref{FtailN} and all similar results
derived below are ``exact'' only in a PN sense. Indeed we have
formally replaced inside the tail integral the orbit of the binary at
any earlier time $t-\tau$ by its orbit at the current time $t$,
thereby neglecting the effect of the binary's adiabatic evolution by
radiation reaction in the past. As a result there should be a
remainder term in~\eqref{FtailN}, given by the order of magnitude of
the adiabatic parameter $\xi_\mathrm{rad}\equiv\dot{\omega}/\omega^2$
associated with the binary's inspiral by radiation reaction. Indeed,
we know~\cite{BD92,BS93} that the replacement of the current motion
inside the tail integral is valid only modulo some remainder
${\cal{O}}\left(\xi_\mathrm{rad}\right)$ or, rather,
${\cal{O}}\left(\xi_\mathrm{rad}\ln\xi_\mathrm{rad}\right)$. In terms
of a PN expansion such remainder brings a correction of relative 2.5PN
order which is always negligible here (indeed the tails are themselves
at 1.5PN order so the total error due the neglect of the influence of
the past in the tails is 4PN).

The other tail integrals, given by the second and third terms in
Eq.~\eqref{Ftail}, are evaluated in exactly the same way. With the PN
accuracy of the present calculation these integrals are truly
Newtonian so the mass octupole moment $I_{ijk}$ and current quadrupole
moment $J_{ij}$ are required at Newtonian order only. For simplicity,
we do not add a superscript (N) to indicate this because there can be
no confusion with other results. We thus need to evaluate the
time-averaged fluxes
\bs\label{Ftailhigher}\bea \langle{\cal{F}}_\mathrm{mass~ oct}\rangle_{\rm tail}
&=& \langle
\frac{4M}{189}\,I_{ijk}^{(4)}(t)\,\int_0^{+\infty}d\tau\,I^{(6)}_{
ijk}(t-\tau)\biggl[\ln\left(\frac{
\tau}{2r_0}\right)+\frac{97}{60}\biggr]\rangle,\\\langle{\cal{F}}_{\rm
curr~ quad}\rangle_{\rm tail} &=& \langle
\frac{64M}{45}\,J_{ij}^{(3)}(t)\,\int_0^{+\infty}d\tau\,J^{(5)}_{
ij}(t-\tau)\biggl[\ln\left(\frac{
\tau}{2r_0}\right)+\frac{7}{6}\biggr]\rangle.  \eea\es
Inserting the Fourier decomposition of the moments, performing the
average using Eq.~\eqref{avN} and using the integration
formula~\eqref{int1} immediately results in
\bs\label{Ftailres}\bea \langle{\cal{F}}_\mathrm{mass~ oct}\rangle_{\rm tail} &=&
\frac{4\pi M}{189}\sum_{p=1}^{+\infty}(p\,n)^{9}\,
\vert\!\mathop{{\cal{I}}}_{(p)}{}_{\!\!ijk}\vert^2,\\
\langle{\cal{F}}_\mathrm{curr~ quad}\rangle_{\rm tail} &=& \frac{64\pi
M}{45}\sum_{p=1}^{+\infty}(p\,n)^{7}\,
\vert\!\mathop{\mathcal{J}}_{(p)}{}_{\!\!ij}\vert^2.  \eea\es
In Sec.~\ref{secV} we shall have to provide some numerical plots for
the eccentricity-dependent enhancement factors associated with
Eqs.~\eqref{Ftailres}, since they cannot be computed analytically.
\subsection{Tails-of-tails and tails squared}
\label{secIVC} 
We have seen that at the 3PN order (\textit{i.e.} 1.5PN beyond the
dominant tail)  the first cubic non-linear interaction, between
the quadrupole moment $I_{ij}$ and two mass monopole factors
$M$, appears. Following Eqs.~\eqref{Ftailtail} we thus have to compute the
``tail-of-tail'' contribution,
\beq\label{avFtailtail} \langle{\cal{F}}_\mathrm{tail(tail)}\rangle =
\langle
\frac{4M^2}{5}\,I_{ij}^{(3)}(t)\,\int_0^{+\infty}d\tau\,I^{(6)}_{
ij}(t-\tau)\biggl[\ln^2\left(\frac{
\tau}{2r_0}\right)+\frac{57}{70}\ln\left(\frac{
\tau}{2r_0}\right)+\frac{124627}{44100}\biggr]\rangle, \eeq
and the so-called ``tail squared'' one,
\beq\label{Ftail2}
\langle{\cal{F}}_{\mathrm{(tail)}^2}\rangle = \langle
\frac{4M^2}{5}\left(\int_0^{+\infty}d\tau\,I^{(5)}_{
ij}(t-\tau)\biggl[\ln\left(\frac{
\tau}{2r_0}\right)+\frac{11}{12}\biggr]\right)^2\rangle.
\eeq
Both contributions are evaluated at relative Newtonian order,
inserting the Fourier decomposition of the Newtonian quadrupole
moment~\eqref{IijfourierN} [suppressing the notation (N) for
simplicity]. The new feature with respect to the previous computation
is the occurrence of a logarithm \textit{squared} in the tail-of-tail
integral~\eqref{avFtailtail}. The integration formula required to deal
with this term is [compare with Eq.~\eqref{int1}]
\beq\label{int2}
\int_0^{+\infty}d\tau\,e^{\ii\,\sigma\,\tau}\ln^2\left(\frac{\tau}{2r_0}\right)
=
\frac{\ii}{\sigma}\left\{\frac{\pi^2}{6}-\left[\frac{\pi}{2}\mathrm{sign}(\sigma)
+ \ii\Bigl(\ln(2\vert \sigma\vert r_0) + C\Bigr)\right]^2\right\}, \eeq
and with this formula, together with~\eqref{int1}, we obtain the
result
\beq\label{Ftailtail2} \langle{\cal{F}}_\mathrm{tail(tail)}\rangle =
\frac{4M^2}{5}\!\sum_{p=1}^{+\infty}(p\,n)^{8}\,
\vert\!\mathop{{\cal{I}}}_{(p)}{}_{\!\!ij}^{\!\!(\mathrm{N})}
\vert^2\left\{\frac{\pi^2}{6}-2\Bigl(\ln(2p\,n \,r_0) +
C\Bigr)^2+\frac{57}{35}\Bigl(\ln(2p\,n\,r_0) +
C\Bigr)-\frac{124627}{22050}\right\}.  \eeq
On the other hand the tail squared term is readily computed
with~\eqref{int1} and found to be
\beq\label{Ftail22} \langle{\cal{F}}_{\mathrm{(tail)}^2}\rangle =
\frac{4M^2}{5}\!\sum_{p=1}^{+\infty}(p\,n)^{8}\,
\vert\!\mathop{{\cal{I}}}_{(p)}{}_{\!\!ij}^{\!\!(\mathrm{N})}
\vert^2\left\{\frac{\pi^2}{2}+2\biggl(\ln(2p\,n \,r_0) + C -
\frac{11}{12}\biggr)^2\right\}.  \eeq
Summing up the two results~\eqref{Ftailtail2} and~\eqref{Ftail22} we
finally obtain
\beq\label{Ftailtailres} \langle{\cal{F}}_{{\rm
tail(tail)+(tail)}^2}\rangle =
\frac{4M^2}{5}\!\sum_{p=1}^{+\infty}(p\,n)^{8}\,
\vert\!\mathop{{\cal{I}}}_{(p)}{}_{\!\!ij}^{\!\!(\mathrm{N})}
\vert^2\left\{\frac{2\pi^2}{3}-\frac{214}{105}\ln(2p\,n\,r_0)
-\frac{214}{105}C-\frac{116761}{29400}\right\}.  \eeq
As we can see the contribution from logarithms \textit{squared} has
cancelled out between the two
terms~\eqref{Ftailtail2}--\eqref{Ftail22}. Such cancellation is in
fact known to occur for general sources~\cite{B98tail}. We observe
also that the result~\eqref{Ftailtailres} still depends on the
arbitrary length scale $r_0$. It will be important to trace out the
fate of this constant and check that the complete energy flux we
obtain at the end (including all the instantaneous contributions
computed in~\cite{ABIQ07}) is independent of $r_0$.

\subsection{The mass quadrupole tail at 1PN order}\label{secIVD}
Let us now tackle the computation of the mass quadrupole tail at the
relative 1PN order, namely
\beq\label{Fquadtail} \langle{\cal{F}}_\mathrm{mass\;quad}\rangle_{\rm tail} =
\langle\frac{4M}{5}\,I_{ij}^{(3)}(t)\,\int_0^{+\infty}d\tau\,I_{ij}^{(5)}(t-\tau)
\biggl[\ln\left(\frac{\tau}{2r_0}\right) +\frac{11}{12}\biggr]\rangle.
\eeq
At the 1PN order (and similarly at any higher PN orders), we must take
care of the doubly-periodic structure of the solution of the motion
[Sec.~\ref{secIIA}], and decompose the multipole moments according to
the general formulas~\eqref{ILJLfourier}. So the 1PN mass quadrupole
moment $I_{ij}$ entering Eq.~\eqref{Fquadtail} is decomposed as
\beq\label{Iijfourier1} I_{ij}(t) =
\sum_{p=-\infty}^{+\infty}\,\sum_{m=-2}^{2}
\,\mathop{{\cal{I}}}_{(p,m)}{}_{\!\!ij}\,e^{\ii\,(p+m\,k)\,\ell}, \eeq
with doubly-indexed Fourier coefficients ${}_{(p,m)}{\cal{I}}_{ij}$
which are valid through order 1PN. We can be more precise and notice
that the harmonics for which $m=\pm 1$ are zero at the 1PN order, so
that
\beq\label{Iijfourier2} I_{ij}(t) =
\sum_{p=-\infty}^{+\infty}\,\left\{
\mathop{{\cal{I}}}_{(p,-2)}{}_{\!\!ij}\,e^{\ii\,(p-2k)\,\ell}
+ \mathop{{\cal{I}}}_{(p,0)}{}_{\!\!ij}\,e^{\ii\,p\,\ell} +
\mathop{{\cal{I}}}_{(p,2)}{}_{\!\!ij}\,e^{\ii\,(p+2 k)\,\ell}\right\},
\eeq
but in the following it is more convenient to work with the general
decomposition~\eqref{Iijfourier1}, keeping in mind that the terms with
$m=\pm 1$ are absent. As before we insert~\eqref{Iijfourier1}
into~\eqref{Fquadtail} to obtain [after neglecting 2.5PN radiation
reaction terms ${\cal{O}}\left(\xi_\mathrm{rad}\right)$]
\bea
\label{Ftailfourier} \langle{\cal{F}}_\mathrm{mass\;quad}\rangle_{\rm tail} &=&
\frac{4M}{5}\!\sum_{p,p';m,m'}n^8(p+m k)^3(p'+m' k)^5
\mathop{{\cal{I}}}_{(p,m)}{}_{\!\!ij}
\mathop{{\cal{I}}}_{(p',m')}{}_{\!\!ij}\nonumber\\&&\quad\quad\times\,\langle
e^{\ii(p+p'+(m+m') k)\ell}\rangle\int_0^{+\infty}d\tau\,e^{-\ii\,(p'+m'
k)\,n \,\tau}\biggl[\ln\left(\frac{
\tau}{2r_0}\right)+\frac{11}{12}\biggr], \eea
where the summations range from $-\infty$ to $+\infty$ for $p$ and
$p'$, and from $-2$ to $2$ for $m$ and $m'$. Evidently the factors
$(p+m k)^3$ and $(p'+m' k)^5$ come from the time-derivatives of the
quadrupole moment. We have explicitly left the last two factors
in~\eqref{Ftailfourier} as they are, namely the average over $\ell$ of
an elementary ``doubly-periodic'' complex exponential, and the Fourier
transform of the tail integral.

The expression~\eqref{Ftailfourier} is to be worked out at the 1PN
order. Since the relativistic advance of the periastron $k$ is already
a small 1PN quantity, the first thing to do is to
evaluate~\eqref{Ftailfourier} at \textit{linear} order in $k$
[\textit{i.e.}, neglecting ${\cal{O}}(k^2)$ which is at least
2PN]. Afterwards we shall insert the explicit expressions for the 1PN
quadrupole moment and ADM mass. We provide here the necessary formulas
for performing the linear-order expansion in $k$ of the last two
factors in~\eqref{Ftailfourier}. The average we perform is over the
orbital period (time to return to the periastron) and so is defined by
\beq\label{avdef}
\langle e^{\ii\,(p+m\,k)\,\ell}\rangle\equiv
\int_0^{2\pi}\frac{d\ell}{2\pi}\,e^{\ii\,(p+m\,k)\,\ell}.
\eeq
Using the fact that $m\,k\ll 1$ since we are in the limit where
$k\rightarrow 0$ (hence $p+m\,k$ is never an integer unless $k=0$), we
readily find
\beq\label{av} \langle e^{\ii\,(p+m\,k)\,\ell}\rangle = \left\{
\begin{array}{ll} \displaystyle \frac{m}{p}\,k & ~~\mathrm{if~} p\neq
0\\[1.5 em] \displaystyle 1+\ii\,\pi \,m \,k & ~~\mathrm{if~}
p=0\end{array}\right\} + {\cal{O}}(k^2).  \eeq
This result depends only on whether $p$ is zero or not, and is true
for any integer $m$, except that when $m=0$ the result~\eqref{av}
becomes ``exact'' as there is no remainder term ${\cal{O}}(k^2)$ in
this case.

On the other hand, to compute the tail integral given by the last
factor in Eq.~\eqref{Ftailfourier}, we expand it at first order in
$k$, obtaining thereby
\beq\label{intk} \int_0^{+\infty}d\tau\,e^{\ii\,(p +m \,k)\,n
\,\tau}\ln\left(\frac{\tau}{2r_0}\right) = \left(1-\frac{m
\,k}{p}\right) \int_0^{+\infty}d\tau\,e^{\ii p\,n
\,\tau}\ln\left(\frac{\tau}{2r_0}\right) - \ii\frac{m \,k}{p^2 n} +
{\cal{O}}(k^2), \eeq
and we apply for the remaining integral in~\eqref{intk} the
formula~\eqref{int1}.

With Eqs.~\eqref{av} and~\eqref{intk} in hand we can explicitly work
out the tail expression~\eqref{Ftailfourier} at first order in $k$
(the extension to higher order in $k$ would in principle be
straightforward). The result will be left in the form of the multiple
Fourier series~\eqref{Ftailfourier}, into which the
results~\eqref{av}--\eqref{intk} have been inserted (we do not try to
give a more explicit form for this result which is given by a
complicated Mathematica expression). In the next Section we shall
re-express this series in terms of some elementary enhancement
functions which will finally be evaluated numerically.
\section{Numerical calculation of the tail integrals}\label{secV}
\subsection{Definition of the eccentricity enhancement factors}\label{secVA}
We define here some functions of the eccentricity by certain Fourier
series of the components of the \textit{Newtonian} multipole moments
$I_L^{(\mathrm{N})}$ and $J_{L-1}^{(\mathrm{N})}$ for a Keplerian
ellipse with eccentricity $e$, semi-major axis $a$, frequency
$n=2\pi/P$ (such that Kepler's law $n^2 a^3=m$ holds at Newtonian
order). In the frame of the center of mass we have
$I_L^{(\mathrm{N})}=\mu s_l(\nu) x^{<L>}$ and $J_{L-1}^{(\mathrm{N})}=\mu
s_l(\nu) x^{<L-2}\varepsilon^{i_{l-1}>ab}x^av^b$ where $\mu=m_1
m_2/m=\nu\,m$. 
Here we pose
$s_l(\nu)\equiv X_2^{l-1}+(-)^lX_1^{l-1}\,,$
 where $\;X_1\equiv\frac{m_1}{m}=\frac{1}{2}\left(1+\sqrt{1-4\nu}\right)\,,$
 and $\;X_2\equiv\frac{m_2}{m}=\frac{1}{2}\left(1-\sqrt{1-4\nu}\right)\,$.
Let us rescale the latter Newtonian moments in order to
make them dimensionless by posing
\bs\bea{I_L^{(\mathrm{N})}} &\equiv& {\mu\,a^l}\, s_l(\nu)\, \hat{I}_L\,, 
\label{ILhat}\\
{J_{L-1}^{(\mathrm{N})}} &\equiv& {\mu\,a^l\,n}\, s_l(\nu)\, \hat{J}_{L-1}\,. \label{JLhat}
\eea\es
Our first ``enhancement'' function is of course the Peters \&
Mathews~\cite{PM63} function which we have already expressed in
Eq.~\eqref{PMf} as a Fourier series [and which turns out to admit the
analytically closed form~\eqref{fe}]. In terms of the Fourier
components of the rescaled quadrupole moment $\hat{I}_{ij}$ this
series reads
\beq
\label{PMf2} f(e) = \frac{1}{16}\sum_{p=1}^{+\infty} p^6
\,\vert\!\mathop{\hat{{\cal{I}}}}_{(p)}{}_{\!\!ij}\vert^2,  \eeq
and is such that the averaged energy flux of compact binaries at the
Newtonian order reads
\beq\label{FN} \langle {\cal{F}}^{(\mathrm{N})} \rangle =
\frac{32}{5}\,\nu^2\,x^5 f(e), \eeq
where we have defined for future convenience the frequency-related PN
parameter $x = (m\,\omega)^{2/3}$ where $\omega$ is the binary's
orbital frequency defined for general orbits by
Eq.~\eqref{omega}. Note that in Eq.~\eqref{FN} which is Newtonian we
can approximate $\omega$ by $n$ (hence $x$ reduces to $m/a$).

\begin{figure}[t]
\centering \includegraphics[scale=1.1]{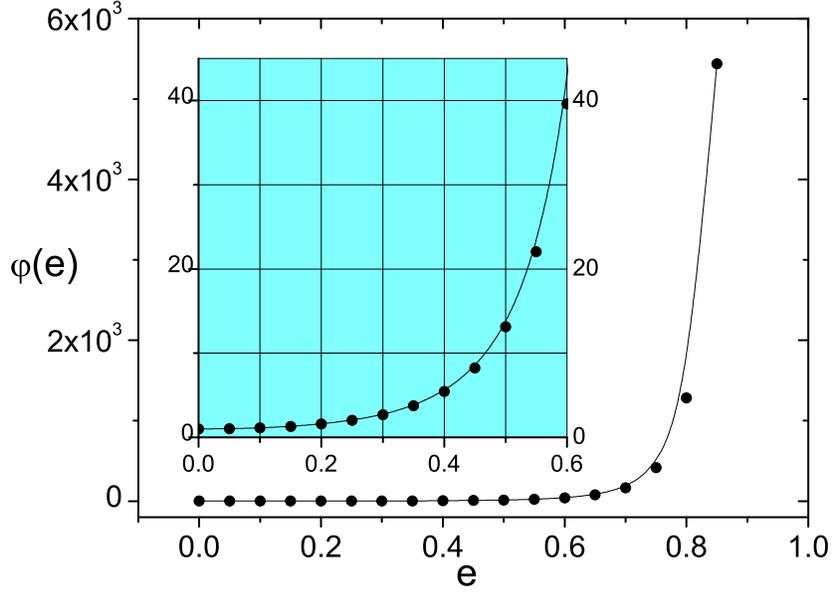}
\caption{
Variation of $\varphi(e)$ with the
eccentricity $e$. The function $\varphi(e)$ agrees with the numerical
calculation of Ref.~\cite{BS93} modulo a trivial rescaling with
$f(e)$. The inset 
graph is a zoom of the function
(which looks like a straight horizontal line in the main graph) at a
smaller scale. The dots represent the numerical computation and the
solid line is a fit to the numerical points. In the circular
orbit limit we have $\varphi(0)=1$.
\label{fig1}}
\end{figure}

Next, we define several other ``enhancement'' functions of the
eccentricity which will permit to usefully parametrize the tail terms
at Newtonian order. First we pose
\beq\label{phie} \varphi(e) = \frac{1}{32}\sum_{p=1}^{+\infty} p^7
\,\vert\!\mathop{\hat{{\cal{I}}}}_{(p)}{}_{\!\!ij}\vert^2.  \eeq
Like for $f(e)$ this function is defined in such a way that it tends
to one in the circular orbit limit, when $e\rightarrow 0$. However,
unlike for $f(e)$, it does not admit a closed-form expression, and
will have to be left in the form of a Fourier series. The function
$\varphi(e)$ parametrizes the mass quadrupole tail at Newtonian order,
in the sense that we have, from Eq.~\eqref{FtailN},
\beq\label{Fphi}
\langle{\cal{F}}_\mathrm{mass\;quad}^{(\mathrm{N})}\rangle =
\frac{32}{5}\,\nu^2\,x^{5}\,\Bigl[ 4\pi\,x^{3/2}\,\varphi(e)\Bigr].
\eeq
For circular orbits, $\varphi(0)=1$ and we recognize the coefficient
$4\pi$ of the 1.5PN tail term ($\propto x^{3/2}$) as computed
numerically in Ref.~\cite{P93} and analytically in
Refs.~\cite{Wi93,BS93}. The function $\varphi(e)$ has already
been computed numerically from its Fourier series~\eqref{phie} in
Ref.~\cite{BS93}. Here we show the plot of $\varphi(e)$ in
Fig.~\ref{fig1} (see Sec.~\ref{secVB} for details on the numerical
computation)\footnote{Note that our notation is different from the one
in~\cite{BS93}; the function $\varphi_\mathrm{BS}(e)$ there is related
to our definition by $\varphi_\mathrm{BS}(e)=\varphi(e)/f(e)$. In the
present work it is better not to divide the various functions by the
Peters \& Mathews function $f(e)$ entering the Newtonian
approximation.}.

\begin{figure}[t]
\centering \includegraphics[width=3in]{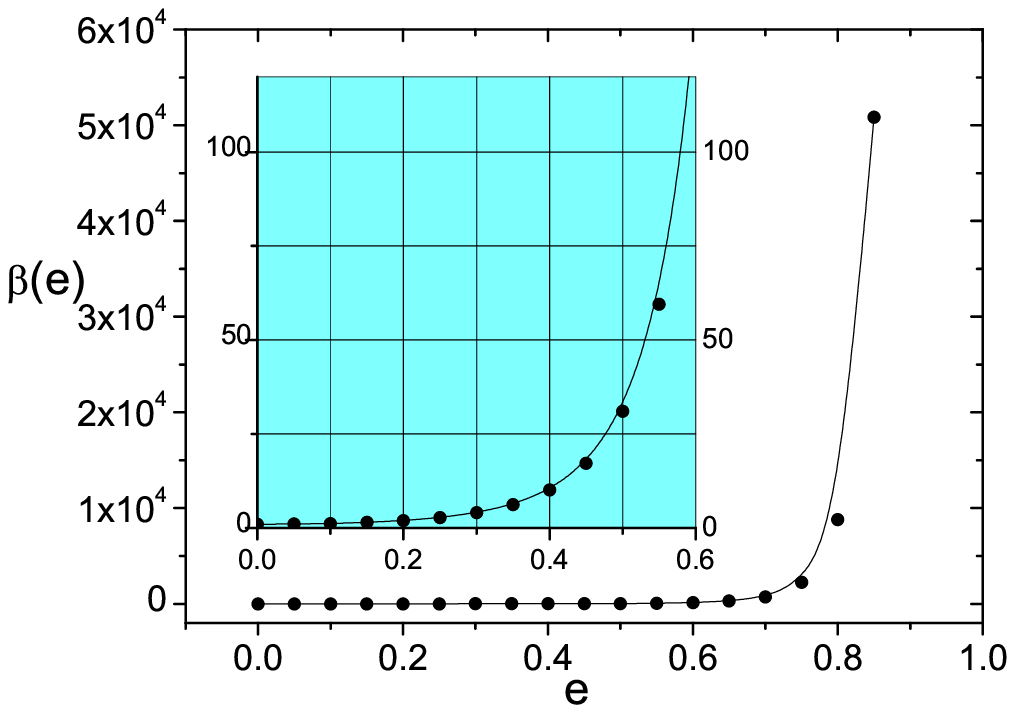}
\includegraphics[width=3in]{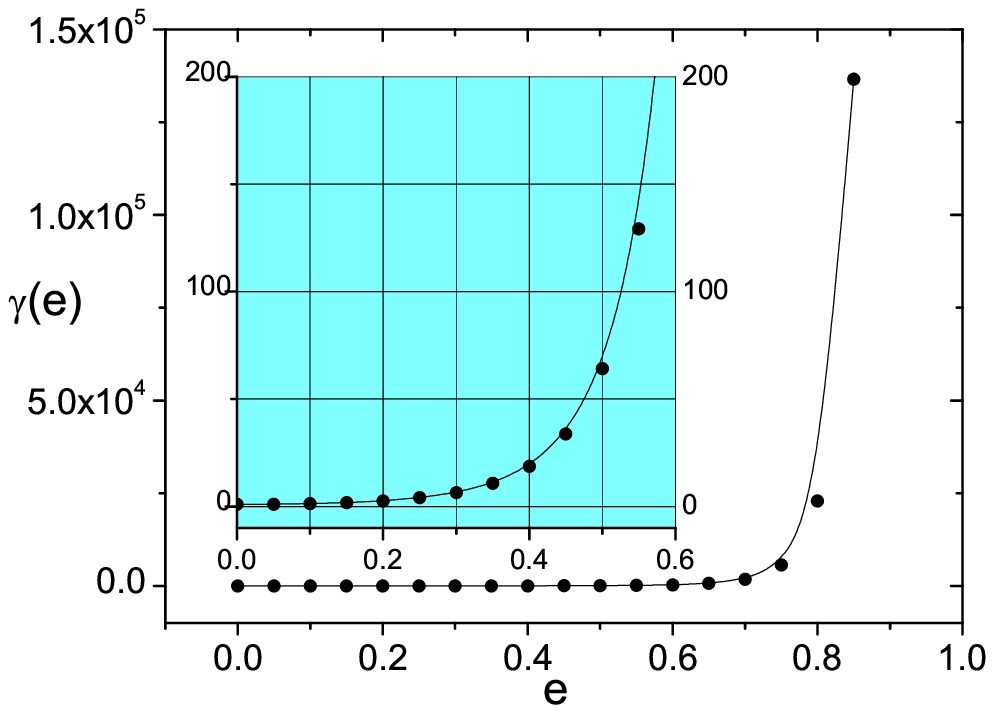}
\caption{
Variation of
$\beta(e)$ (left panel) and $\gamma(e)$ (right panel) with the
eccentricity $e$. In the circular orbit limit we have
$\beta(0)=\gamma(0)=1$.
\label{fig2}}
\end{figure}

We next proceed similarly for the 2.5PN mass octupole and current
quadrupole tails. We pose
\bs \bea \beta(e) &=& \frac{20}{49209}\sum_{p=1}^{+\infty} p^9
\,\vert\!\mathop{\hat{{\cal{I}}}}_{(p)}{}_{\!\!ijk}\vert^2\label{betae},\\
\gamma(e) &=& 4\,\sum_{p=1}^{+\infty}
p^7\,\vert\!\mathop{\hat{\mathcal{J}}}_{(p)}{}_{\!\!ij}\vert^2\label{gammae}.
\eea \es
Again these functions tend to one when $e\rightarrow 0$ (as will be
checked later) and most probably do not admit any closed-form
expressions. With their help these tail terms ($\propto x^{5/2}$) of
Eqs.~\eqref{Ftailhigher} read
\bea \langle{\cal{F}}_\mathrm{mass\; oct}\rangle_{\rm tail} &=&
\frac{32}{5}\,\nu^2\,x^{5}\,\Biggl[\frac{16403}{2016}\,\pi\,(1-4\,\nu)\,
x^{5/2}\,\beta(e)\Biggr],\label{Fbeta}\\ \langle{\cal{F}}_{\rm curr\;
quad}\rangle_{\rm tail} &=&
\frac{32}{5}\,\nu^2\,x^{5}\,\biggl[\frac{\pi}{18}\,(1-4\,\nu)\,x^{5/2}\,
\gamma(e)\biggr].\label{Fgamma} \eea
The numerical graphs of the functions $\beta(e)$ and $\gamma(e)$ are
shown in Fig.~\ref{fig2}.

\begin{figure}[t]
\centering \includegraphics[width=3in]{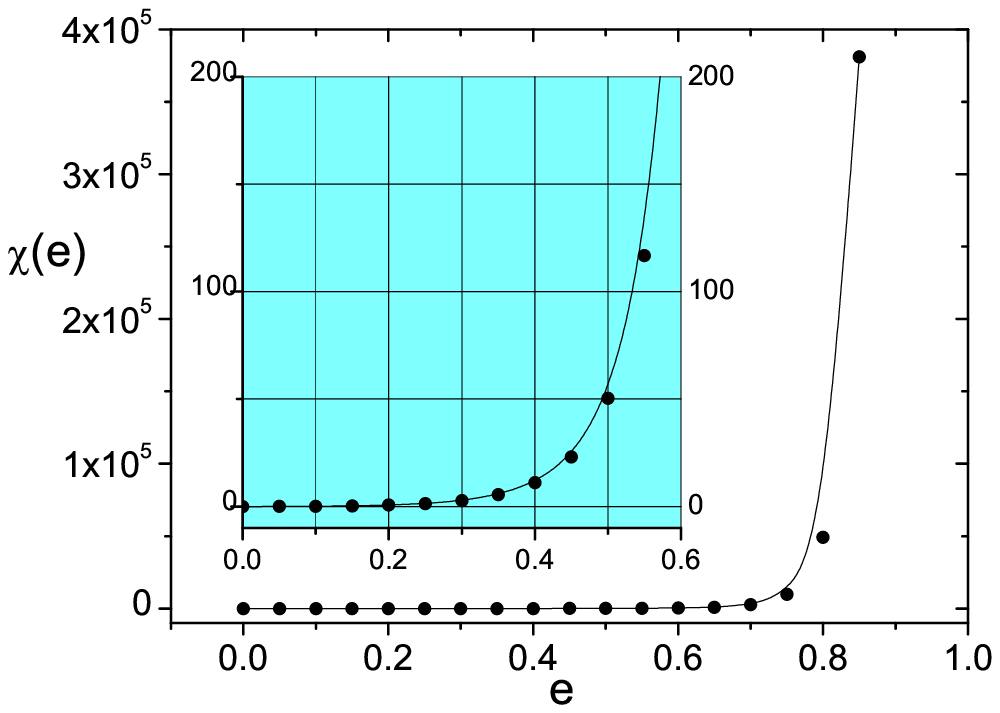}
\includegraphics[width=3in]{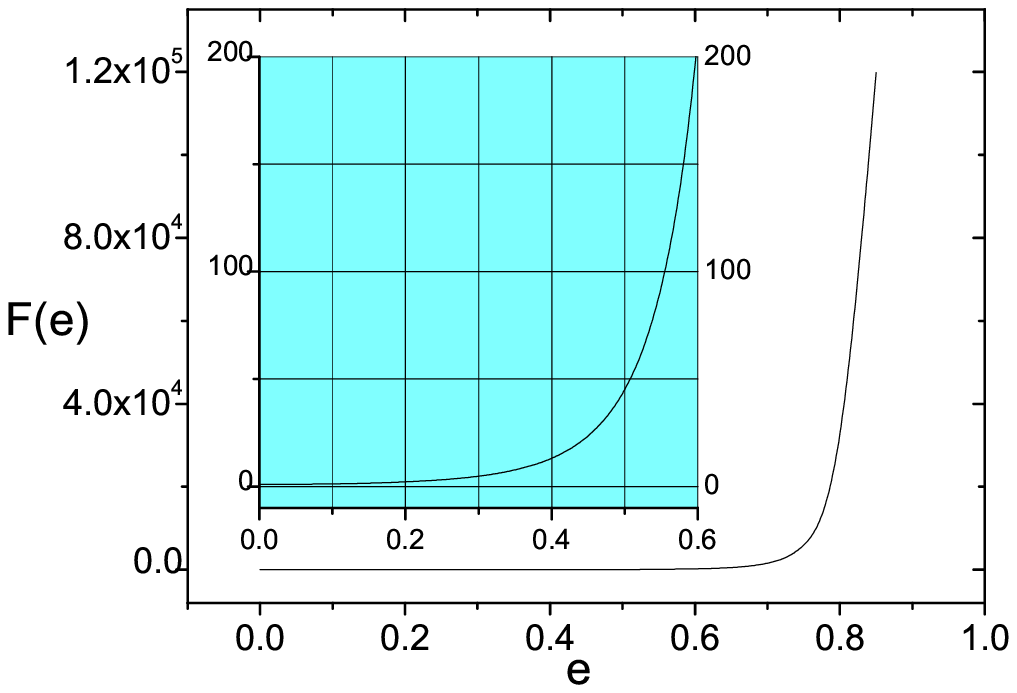}
\caption{
Variation of
$\chi(e)$ (left panel) and $F(e)$ (right panel) with the eccentricity
$e$. In the right panel, the exact expression of $F(e)$ given by
Eq.~\eqref{Fe} is used. In the circular orbit limit we have
$\chi(0)=0$ and $F(0)=1$.
\label{fig3}}
\end{figure}

Two further enhancement factors are then introduced to parametrize
the tail-of-tail and tail squared integrals (which are Newtonian with
the present approximation). The first of these functions looks very
much like the Peters \& Mathews function $f(e)$, Eq.~\eqref{PMf2}, in
the sense that its Fourier series involves \textit{even} powers of the
modes $p$. Namely we define
\beq F(e) = \frac{1}{64}\sum_{p=1}^{+\infty} p^8
\,\vert\!\mathop{\hat{{\cal{I}}}}_{(p)}{}_{\!\!ij}\vert^2\label{Fdef}.
\eeq
Thanks to this even power $\propto p^8$ we find that $F(e)$ can also
be computed as an average performed in the time domain similar to the
one of Eq.~\eqref{feav} for $f(e)$. Namely we easily verify that
\beq\label{Feav}
F(e) = \frac{1}{128\,n^8}
\,\langle{\hat{I}_{ij}^{\,(4)}\hat{I}_{ij}^{\,(4)}}\rangle,
\eeq
which can straightforwardly be computed in the time domain with the
result that $F(e)$ admits like for $f(e)$ an analytic closed form
which is readily obtained as
\beq\label{Fe}
F(e) = \frac{1+\frac{85}{6}e^2+\frac{5171}{192}e^4+\frac{1751}{192}e^6
+\frac{297}{1024}e^8}{(1-e^2)^{13/2}}.
\eeq
On the other hand we shall need to introduce a function whose Fourier
transform differs from the one of $F(e)$ by the presence of the
\textit{logarithm} of modes, namely
\beq \chi(e) = \frac{1}{64}\sum_{p=1}^{+\infty}
p^8\ln\left(\frac{p}{2}\right)
\,\vert\!\mathop{\hat{{\cal{I}}}}_{(p)}{}_{\!\!ij}\vert^2\label{chie}.
\eeq
One can be convinced that very likely $\chi(e)$ does not admit any
analytic form [hence we name it using the Greek alphabet -- in
contrast to $f(e)$ and $F(e)$]. Note that $\chi(e)$ has been
exceptionally defined in such a way that it \textit{vanishes} when
$e\rightarrow 0$. This is easily checked since in the circular orbit
limit (and at Newtonian order) the quadrupole moment
$I_{ij}^{(\mathrm{N})}$ possesses only one harmonic corresponding to  $p=2$
which due to the log term reduces $\chi(e)$ to zero in this case.
 In Fig.~\ref{fig3} we show the numerical plot of
the function $\chi(e)$ [and also the one for $F(e)$].

With those definitions we find that the sum of tail-of-tail and tail
squared contributions obtained in Eq.~\eqref{Ftailtailres} reads
\beq\label{FFPsi}
\langle{\cal{F}}_{\mathrm{tail(tail)+(tail)}^2}\rangle =
\frac{32}{5}\,\nu^2\,x^8 \Biggl\{ \left[ -\frac{116761}{3675} +
\frac{16}{3} \pi^2 -\frac{1712}{105}C -
\frac{1712}{105}\ln\left(4\,\omega\,r_0\right)\right] F(e)
-\frac{1712}{105}\,\chi(e) \Biggr\}.\\ \eeq
The circular-orbit limit can be immediately read off from this expression
and seen to agree with Eq.~(5.9) in Ref.~\cite{B98tail} or Eq.~(12.7)
in Ref.~\cite{BIJ02}.

\begin{figure}[t]
\centering \includegraphics[width=3in]{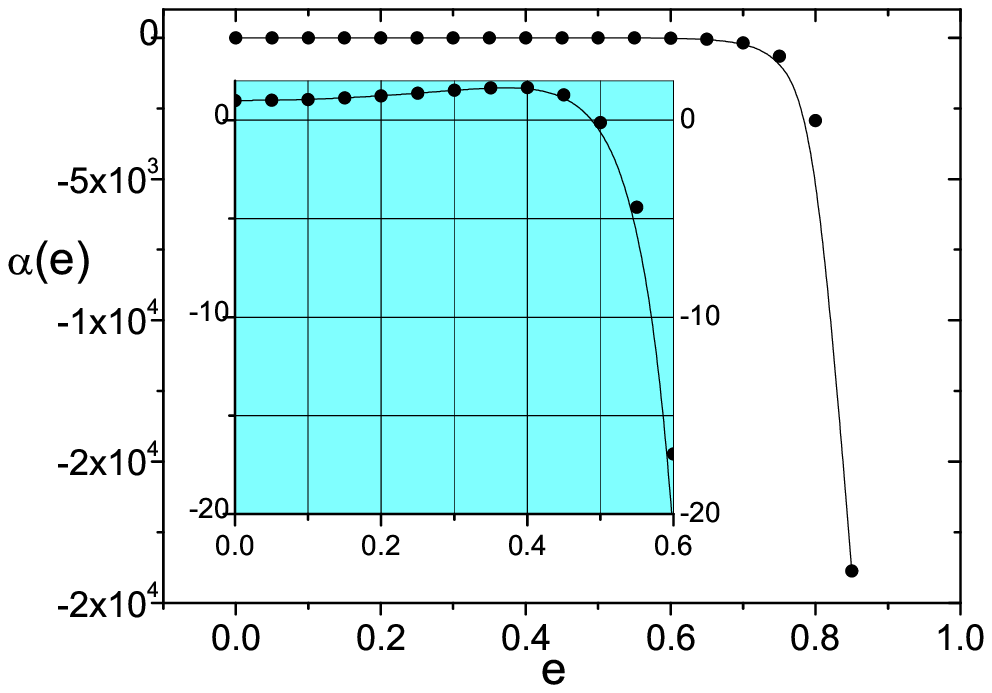}
\includegraphics[width=3in]{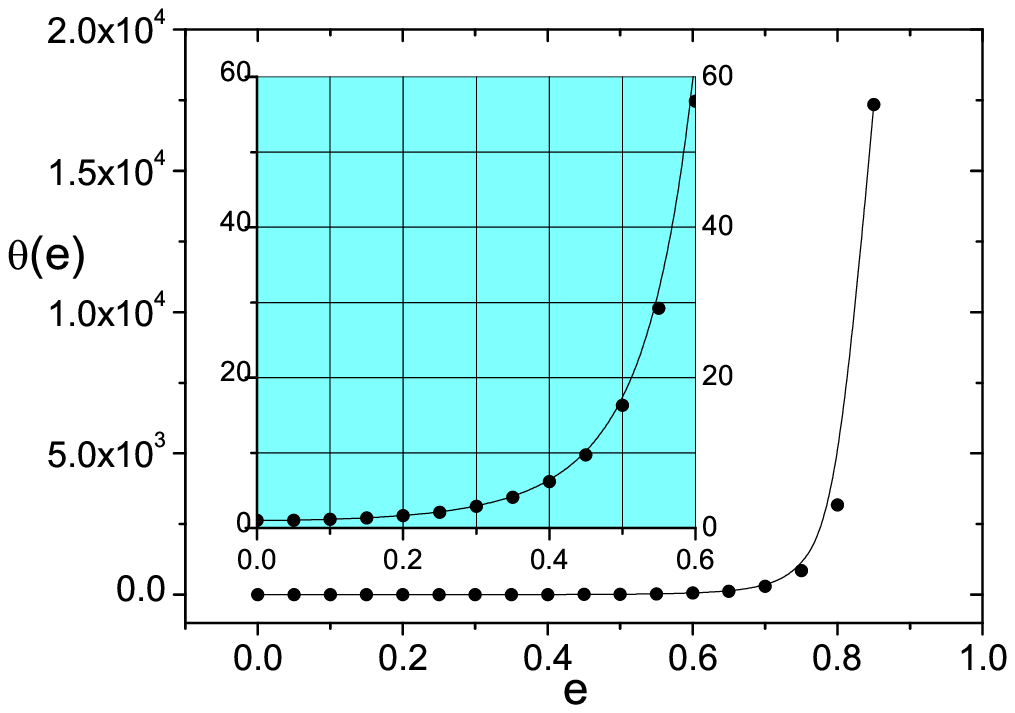}
\caption{
Variation of
$\alpha(e)$ (left panel) and $\theta(e)$ (right panel) with the
eccentricity $e$. In the circular orbit limit we have
$\alpha(0)=\theta(0)=1$.
\label{fig4}}
\end{figure}

Finally we provide the result in the case of the mass quadrupole tail
at 1PN order. We have seen in Sec.~\ref{secIVD} that the calculation
in this case is much more involved, as the Fourier
series~\eqref{Ftailfourier} contains several summations, and depend on
the intermediate results~\eqref{av} and~\eqref{intk}. In addition the
computation must take into account the 1PN relativistic correction in
the mass quadrupole moment and ADM mass; these are provided in
Eqs.~\eqref{Iij1PN} and~\eqref{M1PN} below. We find that probably
there is no simple way [\textit{i.e.} no simple-looking Fourier series
like for instance~\eqref{chie}] for expressing the new enhancement
functions of eccentricity which appear at the 1PN order. However one
can check beforehand that the 1PN term is a linear function of the
symmetric mass ratio $\nu$, hence we must introduce two enhancement
functions, denoted below $\alpha$ and $\theta$. As before we normalize
these functions so that $\alpha(0)=1$ and $\theta(0)=1$. We have
[extending Eq.~\eqref{Fphi} at the 1PN order]
\beq\label{1PNmq} \langle{\cal{F}}_\mathrm{mass\;quad}\rangle_{\rm tail} =
\frac{32}{5}\,\nu^2\,x^{13/2} \Biggl\{
4\pi\,\varphi(e_t)+\pi\,x\,\Biggl[ -\frac{428}{21}\,\alpha(e_t) +
\frac{178}{21} \nu\,\theta(e_t)\Biggr]\Biggr\}.  \eeq
This equation provides the definition of the two enhancement functions
$\alpha$ and $\theta$, and we resort to the Mathematica computation to
obtain them as complicated Fourier decompositions, which will then be
directly computed numerically using the method outlined in
Sec.~\ref{secVB}. Notice that since we are at the 1PN level we must
use a specific definition for the eccentricity, and we adopted here
the ``time'' eccentricity $e_t$ entering the Kepler
equation~\eqref{kepler} in Sec.~\ref{secIIB}. At the 1PN order the
other eccentricities are related to it by Eqs.~\eqref{ratios}. On the
other hand, the frequency-related PN parameter, given by
\beq\label{x} x=(m\,\omega)^{2/3}, \eeq
crucially includes the 1PN relativistic correction coming from the
periastron advance $K=1+k$, through the definition $\omega=n\,K$ of
Sec.~\ref{secIIA}. All the 1PN corrections arising from the
formulas~\eqref{av} and~\eqref{intk}, the multipole moments $M$ and
$I_{ij}$, the use of the time eccentricity $e_t$ and the specific PN
variable $x$, are incorporated in a Mathematica program dealing with
the decomposition~\eqref{Ftailfourier} and used to
obtain~\eqref{1PNmq}. The behaviour of the enhancement functions
$\alpha(e)$ and $\theta(e)$ are given in Fig.~\ref{fig4}.
\subsection{Numerical evaluation of the Fourier coefficients}
\label{secVB}
We now describe the numerical implementation of the procedure for the
computation of the Fourier coefficients of the multipole moments that
lead to the numerical plots of the previous Section. We focus the
discussion on the computation of the crucial coefficients
${}_{(p,m)}\mathcal{I}_{ij}$ at 1PN order which are the more difficult
to obtain. The mass quadrupole moment with 1PN accuracy is given by
[compare with the general structure~\eqref{ILgen1}]
\bea\label{Iij1PN} I_{ij}&=&\mu\,\Biggl\{1+ \left[ v^2\, \left(
\frac{29}{42} - \frac{29 }{14}\nu \right)+\frac{m}{r}\, \left(
-\frac{5}{7}+ \frac{8 }{7}\,\nu \ \right) \right]\, x^{\langle
i}x^{j\rangle}\nonumber\\&&\qquad+\left(\frac{11}{21} -
\frac{11}{7}\nu\right)\,r^2\,v^{\langle i}v^{j\rangle} +
\left(-\frac{4}{7}+\frac{12}{7}\nu\right) r\,\dot{r}\,x^{\langle
i}v^{j\rangle} \Biggr\}, \eea
where $x^i$ and $v^i=d x^i/d t$ are the relative position and velocity
in harmonic coordinates, and $r=\vert x^i\vert$ (like in
Sec.~\ref{secIIB}). Equation~\eqref{Iij1PN} is valid for non-spinning
compact binaries on an arbitrary quasi-Keplerian orbit in the
center-of-mass frame (see \textit{e.g.}~\cite{BI04}). Since we
investigate tails with 1PN relative accuracy we need also the relation
of the ADM mass $M$ to the total mass $m=m_1+m_2$ at 1PN order,
\beq
M=m\left[1+\nu\left(\frac{v^2}{2} -\frac{m}{r}\right) \right].\label{M1PN}
\eeq
Using the quasi-Keplerian representation of the motion
[Sec.~\ref{secIIB}], the dependence of $I_{ij}$ on
$x^i,\,v^i,\,r,\,v^2$ and $\dot r$ can be parametrized in terms of the
eccentric anomaly $u$. However, as explained previously we require
$I_{ij}(\ell)$ in the time domain to proceed. The steps of our
numerical implementation scheme can be summarised as follows:
\begin{enumerate}
\item To begin with, each component of the 1PN mass quadrupole is
expressed in terms of the quasi-Keplerian parameters using
Eqs.~\eqref{rellkepler}--\eqref{V}. The components of the mass
quadrupole are now functions of the eccentric anomaly $u$, and are
parametrized by the mean motion $n$ and by one of the eccentricities
which is chosen to be $e_t$ -- the ``time'' eccentricity in Kepler's
equation~\eqref{kepler}.\,\footnote{The semi-major axis $a_r$ and the
other eccentricities $e_r$ and $e_\phi$ are deduced from $n$ and $e_t$
using Eqs.~\eqref{arn}--\eqref{ratios}.}
\item We next invert, numerically, the equation for the mean anomaly
$\ell=u-e_{t} \sin u$ to obtain the function $u(\ell)$. This can be
done either by using the series representation in terms of Bessel
functions,
\beq
u=\ell+2 \sum_{s=1}^{+\infty}{\frac{1}{s}J_{s}(s\,e_t)\sin(s\,\ell)},
\label{ueqn}
\eeq
or numerically by finding the root of $\ell=u-e_{t} \sin u$. The
latter is a more efficient and more accurate method and we employed it
in this work (we used the FindRoot routine in Mathematica). In this
case we generated a table of 20\,000 points of $u$ and $\ell$ between
$0$ and $2\pi$ (for each value of $e_t$). 
The above inversion enables us to re-express all
functions of the eccentric anomaly $u$ as functions of the mean
anomaly $\ell$. If required, a more accurate implementation for solving
Kepler's equation
along the lines of~\cite{TG07} can be used in the future.
\item One needs to be careful in dealing with the $u$ dependence of
$V$ in Eq.~\eqref{V} to avoid the discontinuity there. To this end it
is best to use
\beq \label{Vu} V(u)= u+ 2\, \arctan \,\biggl(
\frac{\beta_{\phi}\,\sin u}{1-\beta_{\phi}\,\cos u} \biggr) \,, 
\eeq
where $\beta_{\phi}\equiv [1 -(1-e_\phi^2)^{1/2}]/e_\phi$. By this
process, we thus have in hand the Fourier coefficients
${}_{(m)}{\cal{I}}_{ij}(\ell)$ defined in Eq.~\eqref{ILgen3} as
explicit (numerical) functions of $\ell$.
\item Recall that these functions also have a dependence on the mass
ratio $\nu$ and the PN parameter $x$ defined by $(m\,\omega)^{2/3}$
where $\omega=n\,K$. To avoid assuming numerical values for $\nu$ and
$x$ and hence to preserve the full generality of the result, we split
the function ${}_{(m)}\mathcal{I}_{ij}$ into
\beq
\mathop{\mathcal{I}}_{(m)}{}_{\!\!ij}(\ell,e_t,\nu,x) =
\mathop{\mathcal{I}}_{(m)}{}_{\!\!ij}^{\!\!00}(\ell,e_t)+
x\,\left[\mathop{\mathcal{I}}_{(m)}{}_{\!\!ij}^{\!\!\!10}(\ell,e_t)+
\nu\mathop{\mathcal{I}}_{(m)}{}_{\!\!ij}^{\!\!\!11}(\ell,e_t)\right].
\label{calIijexp}
 \eeq
Notice that we have neglected the terms higher than 1PN in writing the
above expression. Now the various ${}_{(m)}\mathcal{I}_{ij}^{\,ab}$
are only functions of $\ell$ and $e_t$. We evaluate the Fourier
coefficients of these terms separately in the next step of the
procedure.
\item For a fixed value of $e_t$, we can straightforwardly get the
plot of ${}_{(m)}\mathcal{I}_{ij}^{\,00}$ versus $\ell$. Equivalently,
one can also write the Fourier decomposition of
${}_{(m)}\mathcal{I}_{ij}^{\,00}(\ell)$ as
\beq \mathop{\mathcal{I}}_{(m)}{}_{\!\!ij}^{\!\!00}(\ell) =
\sum_{p=-\infty}^{+\infty}
\mathop{\mathcal{I}}_{(p,m)}{}_{\!\!ij}^{\!\!\!\!00}\,e^{\ii\,p\,\ell}.
\eeq
Now we seek a numerical fit to Eq.~\eqref{calIijexp}, in powers of
$e^{i p \ell}$, to extract out the coefficients
${}_{(p,m)}\mathcal{I}_{ij}^{\,00}$. We do the same for different
values of $e_t$ and for ${}_{(p,m)}\mathcal{I}_{ij}^{\,10}$ and
${}_{(p,m)}\mathcal{I}_{ij}^{\,11}$.
\item The fitting procedure mentioned above can be implemented either
starting with the STF moment $I_{ij}$ or the non-STF projected
one. The expressions will be different in these two cases as for the
first case the $zz$ component of the moment is not equal to zero by
definition [since $I_{zz}=-(I_{xx}+I_{yy})$] whereas for the latter
case the $zz$ component is zero due to planar motion. This provides a
simple algebraic check on the numerical calculation.
\item Instead of using the basic multipole moment as the starting
function (\textit{e.g.} $I_{ij}$), we find that using the leading time
derivative (\textit{i.e.} $I_{ij}^{(3)}$) improves the numerical
convergence of the results because one deals with lower derivatives of
the basic function. This is very helpful for higher values of
eccentricity.
\item Substituting the Fourier coefficients into
Eq.~\eqref{Ftailfourier} one can generate the numerical values of the
averaged energy flux $\langle{\cal{F}}_\mathrm{mass\;quad}\rangle$ for
the different values of $e_{t}$, and hence get the numerical values of
the enhancement functions, and most importantly of the 1PN ones
$\alpha(e_t)$ and $\theta(e_t)$ defined by~\eqref{1PNmq}. The plots of
these functions reported in Sec.~\ref{secVA} readily follow.
\end{enumerate}

We have just described the procedure for the most difficult 1PN
quadrupole tail yielding the computation of $\alpha(e_t)$ and
$\theta(e_t)$. This procedure is quite general, and provides a method
which could be extended to higher PN orders. 
However, at the Newtonian order it is in fact much more
efficient to make use of the well-known Fourier decomposition of the
Keplerian motion. Using this we can derive the components of the
multipole moments (at Newtonian order) as a series of combinations of
Bessel functions. Then it is a very simple matter to compute
numerically the associated ``Newtonian'' enhancement functions [namely
the functions $\varphi(e)$, $\beta(e)$, $\gamma(e)$ and $\chi(e)$
defined in Sec.~\ref{secVA}]. For the convenience of the reader we
give in Appendix~\ref{appA} all the expressions for each of the components of
the required Newtonian moments [$I_{ij}^{(\mathrm{N})}$,
$I_{ijk}^{(\mathrm{N})}$ and $J_{ij}^{(\mathrm{N})}$] as a series of
Bessel functions. We have used them to compute numerically the enhancement
 functions
$\varphi(e)$, $\beta(e)$, $\gamma(e)$ and $\chi(e)$\footnote{
On the other hand, for the Newtonian tail terms, 
we could proceed exactly in the same way as for the 1PN term, 
following the steps 1 - 8. We have verified that both methods agree well.}.
\section{The hereditary contribution to the 3PN energy flux}\label{secVI}
\subsection{Final expression of the tail terms}\label{secVIA}
\begin{figure}[t]
\centering \includegraphics[width=3in]{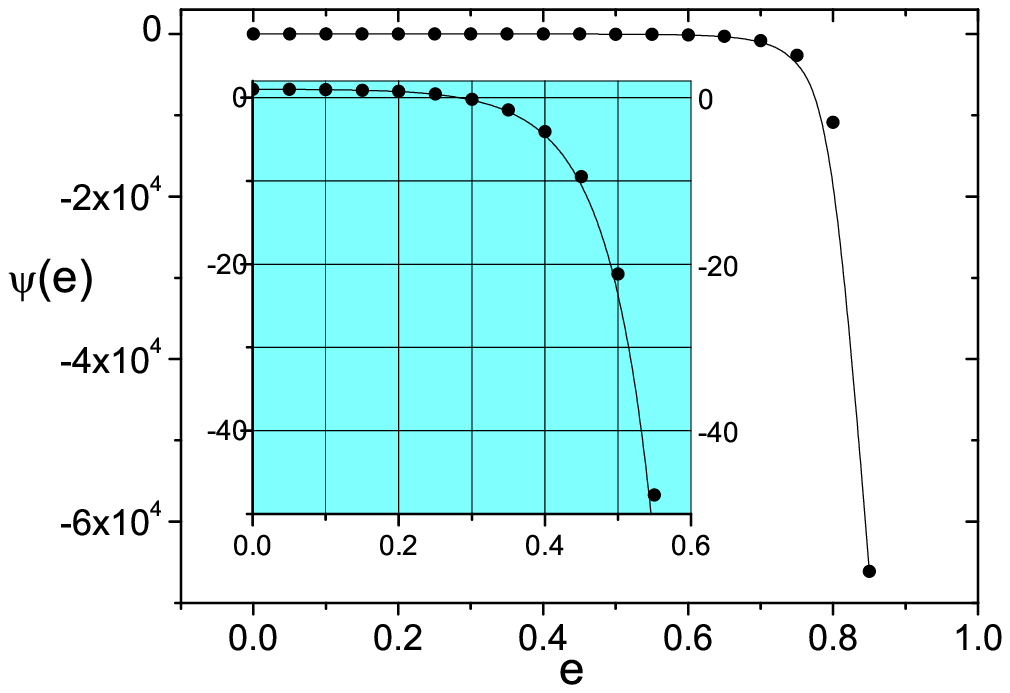}
\includegraphics[width=3in]{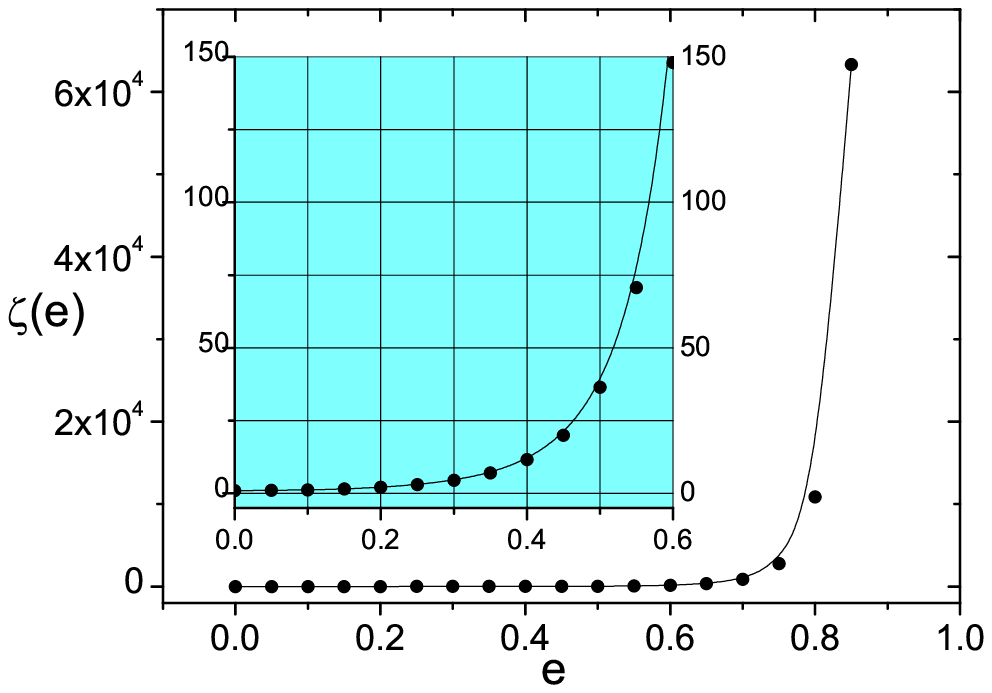}
\caption{
Variation of $\psi(e)$ (left panel) and
$\zeta(e)$ (right panel) with the eccentricity $e$. The 
inset graph is a zoom of the function (which looks like a
straight horizontal line in the main graph) at a smaller scale. The
dots represent the numerical computation and the solid line a
fit to the numerical points. In the circular orbit limit we have
$\psi(0)=\zeta(0)=1$.
\label{fig5}}
\end{figure}

\begin{figure}[t]
\centering \includegraphics[scale=1.1]{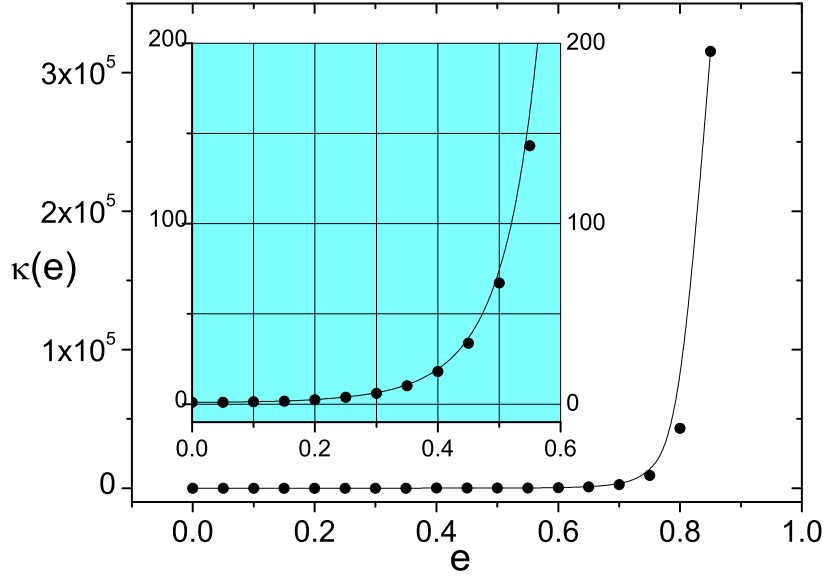}
\caption{
Variation of $\kappa(e)$ with the
eccentricity $e$. In the circular orbit limit we have $\kappa(0)=1$.
\label{fig6}}
\end{figure}

Based on the treatment outlined above of a numerical scheme for the
computation of the orbital average of the hereditary part of the
energy flux up to 3PN, we finally provide the complete results for the
numerical plots of the dimensionless enhancement factors. It is
convenient for the final presentation to redefine in a minor way the
``elementary'' enhancement functions of Sec.~\ref{secVA}, which were
directly given by simple Fourier decompositions. Let us choose
\bs\label{redef}\bea \psi(e) &\equiv& \frac{13696}{8191}\,\alpha(e)
-\frac{16403}{24573}\,\beta(e) - \frac{112}{24573}\,\gamma(e), \\
\zeta(e) &\equiv& -\frac{1424}{4081}\,\theta(e)
+\frac{16403}{12243}\,\beta(e) +\frac{16}{1749}\,\gamma(e), \\
\kappa(e) &\equiv& F(e) +\frac{59920}{116761} \chi(e).  \eea\es
Considering thus the 1.5PN and 2.5PN terms, composed of tails, and the
3PN terms, composed of the tail-of-tail and the tail-squared terms, the
total hereditary contribution to the energy flux~\eqref{Fhered} when
averaged over $\ell$ (and normalized to the Newtonian value for
circular orbits) finally reads
\bea\label{Ftailfinal} \langle{\cal{F}}_\mathrm{hered}\rangle &=&
\frac{32}{5}\,\nu^2\,x^5 \Biggl\{
4\pi\,x^{3/2}\,\varphi(e_t)+\pi\,x^{5/2}
\left[-\frac{8191}{672}\,\psi(e_t)
-\frac{583}{24}\nu\,\zeta(e_t)\right]\nonumber\\ &&+x^3\left[
-\frac{116761}{3675}\,\kappa(e_t) +\left[ \frac{16}{3} \,\pi^2
-\frac{1712}{105}\,C -
\frac{1712}{105}\ln\left(4\omega\,r_0\right)\right]
F(e_t)\right]\Biggr\}.  \eea
In this result all the enhancement functions reduce to one in the
circular case, when $e_t=0$, so the circular-limit is immediately
deduced from inspection of Eq.~\eqref{Ftailfinal}, and is seen to be
in complete agreement with Refs.~\cite{B98tail,BIJ02}. The function
$F(e_t)$ is known analytically, and we recall here its expression,
\beq\label{Fet}
F(e_t) =
\frac{1+\frac{85}{6}e_t^2+\frac{5171}{192}e_t^4+\frac{1751}{192}e_t^6
+\frac{297}{1024}e_t^8}{(1-e_t^2)^{13/2}}.
\eeq
However the other enhancement functions $\varphi(e_t)$, $\psi(e_t)$,
$\zeta(e_t)$ and $\kappa(e_t)$ in Eq.~\eqref{Ftailfinal} (very likely)
do not admit any analytic closed-form expressions.  We have explained
in Sec.~\ref{secVB} the details of the numerical calculation of these
functions. We now present the numerical plots of the final functions
$\psi(e_t)$, $\zeta(e_t)$ and $\kappa(e_t)$ in
Figs.~\ref{fig5}--\ref{fig6} as functions of the eccentricity $e_t$
[recall that the function $\varphi(e_t)$ has already been given in
Fig.~\ref{fig1}]~\footnote{The numerical results used for the figures 1-6 are
available in the form of Tables on request from the authors.}.

As seen from Eq.~\eqref{Ftailfinal} the final result depends on the
constant $r_0$ at the 3PN order. Let us understand in bit more detail
the occurrence of this constant. We first recall from
Ref.~\cite{B98tail} that the dependence on the constant $r_0$ of the
\textit{radiative} quadrupole moment at infinity, say $U_{ij}$, arises
precisely at the 3PN order, and comes exclusively from the
contribution of tails-of-tails (\textit{i.e.} the cubic multipole
interaction $M^2\times I_{ij}$). It is explicitly given by
\beq\label{Uij}
U_{ij}(t) = I_{ij}^{(2)}(t) + \cdots +\frac{214}{105}
\,M^2\,I_{ij}^{(4)}(t)\,\ln r_0 + \cdots,
\eeq
in which we have indicated that $U_{ij}$ simply reduces to the second
time derivative of $I_{ij}$ at the Newtonian order, and where we show
the only term which depends on the constant $r_0$; such a term appears
at 3PN order and turns out to be proportional to the fourth time
derivative of $I_{ij}$. The dots in Eq.~\eqref{Uij} denote many terms
which do not depend on $r_0$. From~\eqref{Uij} it is then trivial to
deduce that the corresponding dependence on $r_0$ of the averaged
energy flux at 3PN order must be
\bea\langle{\cal{F}}^{\mathrm{(3PN)}}\rangle &=& \frac{1}{5}
\langle{U_{ij}^{(1)}U_{ij}^{(1)}}\rangle + \cdots\nonumber\\ &=& \frac{1}{5}
\langle{I_{ij}^{(3)}I_{ij}^{(3)}}\rangle + \cdots + \frac{428}{525}
\,M^2\,\langle{I_{ij}^{(3)}I_{ij}^{(5)}}\rangle\,\ln r_0 +
\cdots.\label{Ftailr00} \eea
Now we can take advantage of the fact that inside the operation of
averaging over $\ell$ [denoted by $\langle\rangle$ and defined
by~\eqref{avdef}] one can freely operate by parts the time
derivatives. Hence, we can write that
$\langle{I_{ij}^{(3)}I_{ij}^{(5)}}\rangle=-\langle{I_{ij}^{(4)}I_{ij}^{(4)}}\rangle$
and so we arrive at the result
\beq\label{Ftailr0} \langle{\cal{F}}^{\mathrm{(3PN)}}\rangle =
\frac{1}{5} \langle{I_{ij}^{(3)}I_{ij}^{(3)}}\rangle + \cdots -
\frac{428}{525} \,M^2\,\langle{I_{ij}^{(4)}I_{ij}^{(4)}}\rangle\,\ln
r_0 + \cdots. \eeq
The factor of $\ln r_0$ in Eq.~\eqref{Ftailr0} looks like a
``quadrupole formula'' but where the third time derivative of the
moment would be replaced by the fourth one. Notice that the above
expression has been computed for general radiative-type moments and is
true for any PN source, in particular for a binary system moving on an
eccentric orbit. Therefore the dependence on $\ln r_0$ found
in~\eqref{Ftailr0} should perfectly match with the one we have
obtained in Eq.~\eqref{Ftailfinal}. Thus, comparing
with~\eqref{Ftailfinal}, one readily infers that the function $F(e_t)$
in the case of an eccentric binary must necessarily be given by the
components of the quadrupole moment in the time domain as
\beq\label{Fe0} F(e_t) = \frac{M^2}{128\,\nu^2\,x^8}
\,\langle{I_{ij}^{(4)}I_{ij}^{(4)}}\rangle .  \eeq
This prediction is perfectly in agreement with our finding for the
function $F(e_t)$ in Eq.~\eqref{Feav} (indeed, since we are at leading
order, $M$ reduces to $m$, $e_t$ agrees with $e$, $\omega$ equals
$n$). We have therefore confirmed the correctness of the dependence
upon $r_0$ of Eq.~\eqref{Ftailfinal}.

We already know from the study of the circular-orbit case
(\textit{cf.}~\cite{BIJ02}) that the dependence on $r_0$ is cancelled
out with a similar term contained in the expression of the source-type
quadrupole moment $I_{ij}$ at 3PN order. This cancellation must in
fact be true for general sources, and has been proved on general
grounds in Ref.~\cite{B98tail}. It will therefore give an interesting
check of our calculations when we show in the companion
paper~\cite{ABIQ07} that the cancellation of $r_0$ occurs for general
eccentric orbits.

To finish let us provide here the expressions of our final enhancement
functions at the first order in $e_t^2$ when $e_t\rightarrow 0$. These
expansions will be useful in the following paper~\cite{ABIQ07}, when
we compare the perturbative limit of the complete energy flux at 3PN
order (including all instantaneous terms) with the result of
black-hole perturbations. Note that those expansions are obtained
analytically. For the functions which are Newtonian we can
either use the Fourier
coefficients in the Appendix~\ref{appA} and expand them at first order
in $e_t^2$ or follow the general procedure explained in
Sec.~\ref{secVB}
 for the relevant moments but expanding  Eq.~\eqref{ueqn} to only
first order in $e_t^2$, namely,
\beq
u=\ell +e_t\sin \ell +\frac{e_t^2}{2}\sin 2\ell +{\cal{O}}\left(e_t^3\right).
\eeq 
Concerning the two 1PN functions [$\psi(e_t)$ and
$\zeta(e_t)$], on the other hand,
 we obtain them directly using the latter
procedure. 
We find 
\bs\label{enhancexp}
\bea
\label{phiexp} \varphi\left(e_t\right)&=& 1+\frac{2335
}{192}\,e_t^2+\mathcal{O}\left(e_t^4\right),\\
\label{psiexp} \psi\left(e_t\right)&=& 1-\frac{22988
}{8191}\,\,e_t^2+\mathcal{O}\left(e_t^4\right),\\
\label{zetaexp} \zeta\left(e_t\right)&=& 1+
\frac{1011565} {48972}  \,e_t^2+\mathcal{O}\left(e_t^4\right),\\
\label{kappaexp} \kappa\left(e_t\right)&=& 1+\left(\frac{62}{3}
-\frac{4613840}{350283}\ln 2+\frac{24570945}{1868176}\ln 3\right)\,e_t^2
+\mathcal{O}\left(e_t^4\right), \eea \es
and of course [since this is immediately deduced from Eq.~\eqref{Fet}]
\beq\label{Fetexp}
F\left(e_t\right) = 1+\frac{62}{3}\,e_t^2
+\mathcal{O}\left(e_t^4\right).
\eeq
We have checked that the numerical results of Figs.\ref{fig1},
\ref{fig5} and \ref{fig6} agree well with Eqs.~\eqref{enhancexp} in the
limit of small eccentricities.
\subsection{Conclusion and future directions}\label{secVIB}
The far-zone flux of energy contains hereditary contributions that
depend on the entire past history of the source. Using the GW
generation formalism consisting of a multipolar post-Minkowskian
expansion with matching to a PN source, we have proposed and
implemented a semi-analytical method to compute the hereditary
contributions from the inspiral phase of a binary system of compact
objects moving on quasi-elliptical orbits up to 3PN order.  The method
explicitly uses the 1PN quasi-Keplerian representation of elliptical
orbits and exploits the  doubly periodic nature of the motion to
average the fluxes over the binary's orbit. Together with the
instantaneous contributions evaluated in the next paper~\cite{ABIQ07},
it provides crucial inputs for the construction of ready-to-use
templates for binaries moving on eccentric orbits, an interesting
class of sources for the ground based gravitational wave detectors
LIGO/Virgo and especially space based detectors like LISA.

The extension of these methods to compute the hereditary terms in the
3PN angular momentum flux and 2PN linear momentum flux is the next
step required to proceed towards the above goal and is currently under investigation.
 The extension to compute the 3.5PN terms for elliptical orbits
is currently not possible due to some as yet uncalculated terms in
  the generation
formalism at this order for general orbits. It would also require the
use of the 2PN generalised quasi-Keplerian representation for some of
the leading multipole moments.

\medskip

\acknowledgments L.B. and B.R.I. thank the Indo-French Collaboration
(IFCPAR) under which this work has been carried out.
M.S.S.Q. acknowledges the Indo-Yemen cultural exchange programme.
B.R.I. acknowledges the
hospitality of the Institut Henri Poincar\'e and Institut des Hautes
Etudes Scientifiques during the final stages of the writing of the
paper.
Most of algebraic calculations leading to the results of this paper
are done with the software Mathematica.  

\appendix
\section{Fourier coefficients of the multipole moments}\label{appA}
In this Appendix we provide the expressions of the Fourier
coefficients of the \textit{Newtonian} multipole moments in terms of
combinations of Bessel functions. We decompose the components of the
moments as Fourier series,
\bs\bea I_L^{(\mathrm{N})}(t) &=&
\sum_{p=-\infty}^{+\infty}\,\mathop{{\cal{I}}}_{(p)}{}_{\!\!L}^{(\mathrm{N})}\,e^{\ii
p \ell},\\ J_{L-1}^{(\mathrm{N})}(t) &=&
\sum_{p=-\infty}^{+\infty}\,\mathop{\mathcal{J}}_{(p)}{}_{\!\!\!L-1}^{(\mathrm{N})}\,e^{\ii
p\ell},\eea\es
where the Fourier coefficients can be obtained by evaluating the
following integrals
\bs\bea \mathop{{\cal{I}}}_{(p)}{}_{\!\!L}^{(\mathrm{N})}&=&
\frac{1}{2\pi}\int_0^{2\pi}d\ell\,I_L^{(\mathrm{N})}(t) \,e^{-\ii p
\ell},\\ \mathop{\mathcal{J}}_{(p)}{}_{\!\!\!L-1}^{(\mathrm{N})}&=&
\frac{1}{2\pi}\int_0^{2\pi}d\ell\,J_{L-1}^{(\mathrm{N})}(t) \,e^{-\ii p
\ell}.  \eea\es
For the mass quadrupole moment at Newtonian order  we have\footnote{Note that the Fourier coefficients
we provide are for {\it normalized} multipole moments
as defined in Eqs~(\ref{ILhat})--(\ref{JLhat}).}
\bs
\label{A3}
\bea
\mathop{{\cal{I}}}_{(p)}{}_{\!\!xx}^{(\mathrm{N})}&=&{\left(\frac{1}{6}+\frac{3}{2}
e_t^2\right) J_p\left(p e_t\right)}\nonumber\\&& {+\left(-\frac{7}{8}
e_t-\frac{3}{8} e_t^3\right) \left(J_{p-1}\left(p
e_t\right)+J_{p+1}\left(p e_t\right)\right)}\nonumber\\&&
{+\left(\frac{1}{4}+\frac{1}{4} e_t^2\right) \left(J_{p-2}\left(p
e_t\right)+J_{p+2}\left(p e_t\right)\right)}\nonumber\\&&
{+\left(-\frac{1}{8} e_t+\frac{1}{24} e_t^3\right)
\left(J_{p-3}\left(p e_t\right)+J_{p+3}\left(p e_t\right)\right)},\\
\mathop{{\cal{I}}}_{(p)}{}_{\!\!xy}^{(\mathrm{N})}&=&{-\ii
\sqrt{1-e_t^2}\left\{\frac{5}{8} e_t \left(-J_{p-1}\left(p
e_t\right)+J_{p+1}\left(p e_t\right)\right)\right.}\nonumber\\&&
{+\left(-\frac{1}{4}-\frac{1}{4} e_t^2\right) \left(J_{p+2}\left(p
e_t\right)-J_{p-2}\left(p e_t\right)\right)}\nonumber\\&&
{\left.+\frac{1}{8} e_t \left(J_{p+3}\left(p
e_t\right)-J_{p-3}\left(p e_t\right)\right)\right\}},\\
\mathop{{\cal{I}}}_{(p)}{}_{\!\!yy}^{(\mathrm{N})}&=&{\left(\frac{1}{6}-e_t^2\right)
J_p\left(p e_t\right)}\nonumber\\&& {+\left(\frac{3}{8}
e_t+\frac{1}{4} e_t^3\right) \left(J_{p-1}\left(p
e_t\right)+J_{p+1}\left(p e_t\right)\right)}\nonumber\\&&
{-\frac{1}{4} \left(J_{p-2}\left(p e_t\right)+J_{p+2}\left(p
e_t\right)\right)}\nonumber\\&& {+\left(\frac{1}{8} e_t-\frac{1}{12}
e_t^3\right)
\left(J_{p-3}\left(p e_t\right)+J_{p+3}\left(p e_t\right)\right)},\\
\mathop{{\cal{I}}}_{(p)}{}_{\!\!zz}^{(\mathrm{N})}&=&{\left(-\frac{1}{3}-\frac{1}{2}
e_t^2\right) J_p\left(p e_t\right)}\nonumber\\&& {+\left(\frac{1}{2}
e_t+\frac{1}{8} e_t^3\right) \left(J_{p-1}\left(p
e_t\right)+J_{p+1}\left(p e_t\right)\right)}\nonumber\\&&
{-\frac{1}{4} e_t^2 \left(J_{p-2}\left(p e_t\right)+J_{p+2}\left(p
e_t\right)\right)}\nonumber\\&& {+\frac{1}{24} e_t^3
\left(J_{p-3}\left(p e_t\right)+J_{p+3}\left(p e_t\right)\right)}.
\eea\es
For the mass octupole moment we find
\bs\bea
\mathop{{\cal{I}}}_{(p)}{}_{\!\!xxx}^{(\mathrm{N})}&=&{-\left\{\left(\frac{3}{8}
e_t+\frac{11}{8} e_t^3\right) J_p\left(p e_t\right)\right.}\nonumber\\&&
{+\left(-\frac{3}{40}-\frac{21}{20} e_t^2-\frac{11}{40} e_t^4\right)
\left(J_{p-1}\left(p e_t\right)+J_{p+1}\left(p
e_t\right)\right)}\nonumber\\&& {+\left(\frac{11}{20}
e_t+\frac{3}{20} e_t^3\right) \left(J_{p-2}\left(p
e_t\right)+J_{p+2}\left(p e_t\right)\right)}\nonumber\\&&
{+\left(-\frac{1}{8}-\frac{3}{20} e_t^2+\frac{3}{40} e_t^4\right)
\left(J_{p-3}\left(p e_t\right)+J_{p+3}\left(p
e_t\right)\right)}\nonumber\\&& {+\left.\left(\frac{1}{16}
e_t-\frac{3}{80} e_t^3\right) \left(J_{p-4}\left(p
e_t\right)+J_{p+4}\left(p e_t\right)\right)\right\}},\\
\mathop{{\cal{I}}}_{(p)}{}_{\!\!xxy}^{(\mathrm{N})}&=&{\ii \sqrt{1-e_t^2}
\left\{\left(\frac{1}{40}+\frac{27}{40} e_t^2\right)
\left(J_{p+1}\left(p e_t\right)-J_{p-1}\left(p
e_t\right)\right)\right.}\nonumber\\&& {+\left(-\frac{19}{40}
e_t-\frac{9}{40} e_t^3\right) \left(J_{p+2}\left(p
e_t\right)-J_{p-2}\left(p e_t\right)\right)}\nonumber\\&&
{+\left(\frac{1}{8}+\frac{7}{40} e_t^2\right) \left(J_{p+3}\left(p
e_t\right)-J_{p-3}\left(p e_t\right)\right)}\nonumber\\&&
{\left.+\left(-\frac{1}{16} e_t+\frac{1}{80} e_t^3\right)
\left(J_{p+4}\left(p e_t\right)-J_{p-4}\left(p
e_t\right)\right)\right\}},\\
\mathop{{\cal{I}}}_{(p)}{}_{\!\!xyy}^{(\mathrm{N})}&=&{-\left\{\left(\frac{1}{8}
e_t- e_t^3\right) J_p\left(p e_t\right)\right.}\nonumber\\&&
{+\left(-\frac{1}{40}+\frac{21}{40} e_t^2+\frac{1}{5} e_t^4\right)
\left(J_{p-1}\left(p e_t\right)+J_{p+1}\left(p
e_t\right)\right)}\nonumber\\&& {+\left(-\frac{2}{5}
e_t+\frac{1}{20} e_t^3\right) \left(J_{p-2}\left(p
e_t\right)+J_{p+2}\left(p e_t\right)\right)}\nonumber\\&&
{+\left(\frac{1}{8}+\frac{3}{40} e_t^2-\frac{1}{10} e_t^4\right)
\left(J_{p-3}\left(p e_t\right)+J_{p+3}\left(p
e_t\right)\right)}\nonumber\\&& {\left.+\left(-\frac{1}{16}
e_t+\frac{1}{20} e_t^3\right) \left(J_{p-4}\left(p
e_t\right)+J_{p+4}\left(p e_t\right)\right)\right\}},\\
\mathop{{\cal{I}}}_{(p)}{}_{\!\!yyy}^{(\mathrm{N})}&=&{\ii \sqrt{1-e_t^2}
\left\{\left(\frac{3}{40}-\frac{3}{5} e_t^2\right)
\left(-J_{p-1}\left(p e_t\right)+J_{p+1}\left(p
e_t\right)\right)\right.}\nonumber\\&& {+\left(\frac{13}{40}
e_t+\frac{1}{5} e_t^3\right) \left(-J_{p-2}\left(p
e_t\right)+J_{p+2}\left(p e_t\right)\right)}\nonumber\\&&
{+\left(-\frac{1}{8}-\frac{1}{10} e_t^2\right) \left(-J_{p-3}\left(p
e_t\right)+J_{p+3}\left(p e_t\right)\right)}\nonumber\\&&
{\left.+\left(\frac{1}{16} e_t-\frac{1}{40} e_t^3\right)
\left(-J_{p-4}\left(p e_t\right)+J_{p+4}\left(p
e_t\right)\right)\right\}},\\
\mathop{{\cal{I}}}_{(p)}{}_{\!\!zzx}^{(\mathrm{N})}&=&{-\left\{\left(-\frac{1}{2}
e_t-\frac{3}{8} e_t^3\right) J_p\left(p e_t\right)\right.}\nonumber\\&&
{+\left(\frac{1}{10}+\frac{21}{40} e_t^2+\frac{3}{40} e_t^4\right)
\left(J_{p-1}\left(p e_t\right)+J_{p+1}\left(p
e_t\right)\right)}\nonumber\\&& {+\left(-\frac{3}{20}
e_t-\frac{1}{5} e_t^3\right) \left(J_{p-2}\left(p
e_t\right)+J_{p+2}\left(p e_t\right)\right)}\nonumber\\&&
{+\left(\frac{3}{40} e_t^2+\frac{1}{40} e_t^4\right)
\left(J_{p-3}\left(p e_t\right)+J_{p+3}\left(p
e_t\right)\right)}\nonumber\\&& {\left.-\frac{1}{80} e_t^3
\left(J_{p-4}\left(p e_t\right)+J_{p+4}\left(p e_t\right)\right)\right\}},\\
\mathop{{\cal{I}}}_{(p)}{}_{\!\!zzy}^{(\mathrm{N})}&=&{\ii \sqrt{1-e_t^2}
\left\{\left(-\frac{1}{10}-\frac{3}{40} e_t^2\right)
\left(-J_{p-1}\left(p e_t\right)+J_{p+1}\left(p
e_t\right)\right)\right.}\nonumber\\&& {+\left(\frac{3}{20}
e_t+\frac{1}{40} e_t^3\right) \left(-J_{p-2}\left(p
e_t\right)+J_{p+2}\left(p e_t\right)\right)}\nonumber\\&&
{-\frac{3}{40} e_t^2 \left(-J_{p-3}\left(p e_t\right)+J_{p+3}\left(p
e_t\right)\right)}\nonumber\\&& {\left.+\frac{1}{80} e_t^3
\left(-J_{p-4}\left(p e_t\right)+J_{p+4}\left(p
e_t\right)\right)\right\}}.
.\eea\es
Finally, for the current quadrupole moment,
\bs
\bea
\mathop{\mathcal{J}}_{(p)}{}_{\!\!xz}^{(\mathrm{N})}&=&{-\frac{1}{4} \sqrt{1-e_t^2}
\left\{3 e_t J_p\left(p e_t\right)\right.}\nonumber\\&&
{-\frac{1}{4}\left(1+e_t^2\right) \left(J_{p-1}\left(p
e_t\right)+J_{p+1}\left(p e_t\right)\right)}\nonumber\\&&
{\left.+\frac{1}{8} e_t \left(J_{p-2}\left(p
e_t\right)+J_{p+2}\left(p e_t\right)\right)\right\}},\\
\mathop{\mathcal{J}}_{(p)}{}_{\!\!yz}^{(\mathrm{N})}&=&{\frac{\ii}{4}
\left(1-e_t^2\right)\left\{\left(J_{p+1}\left(p
e_t\right)-J_{p-1}\left(p e_t\right)\right)\right.}\nonumber\\&&
{\left.-\frac{1}{2} e_t \left(J_{p+2}\left(p
e_t\right)-J_{p-2}\left(p e_t\right)\right)\right\}}.
\eea\es
\bibliography{/home/arun/tphome/arun/ref-list}
\end{document}